\documentclass[aps, prx, twocolumn, superscriptaddress, longbibliography]{revtex4-1}
\usepackage{graphicx}
\usepackage{dcolumn}
\usepackage{bm}
\usepackage{amssymb,amsfonts,amsmath}
\usepackage{multirow}
\usepackage{braket}
\usepackage{mathtools}
\usepackage{array}
\usepackage{hyperref}

\usepackage{color}

\begin{document}
\title{Benchmarking Gate Fidelities in a Si/SiGe Two-Qubit Device}

\author{X.~Xue}
\affiliation{QuTech, Delft University of Technology, Lorentzweg 1, 2628 CJ Delft, The Netherlands}
\affiliation{Kavli Institute of Nanosicence, Delft University of Technology, 2600 GA Delft, The Netherlands}
\author{T.~F.~Watson}
\affiliation{QuTech, Delft University of Technology, Lorentzweg 1, 2628 CJ Delft, The Netherlands}
\affiliation{Kavli Institute of Nanosicence, Delft University of Technology, 2600 GA Delft, The Netherlands}
\author{J.~Helsen}
\affiliation{QuTech, Delft University of Technology, Lorentzweg 1, 2628 CJ Delft, The Netherlands}
\author{D.~R.~Ward}
\affiliation{University of Wisconsin-Madison, Madison, WI 53706, USA}
\author{D.~E.~Savage}
\affiliation{University of Wisconsin-Madison, Madison, WI 53706, USA}
\author{M.~G.~Lagally}
\affiliation{University of Wisconsin-Madison, Madison, WI 53706, USA}
\author{S.~N.~Coppersmith}
\affiliation{University of Wisconsin-Madison, Madison, WI 53706, USA}
\author{M.~A.~Eriksson}
\affiliation{University of Wisconsin-Madison, Madison, WI 53706, USA}
\author{S.~Wehner}
\affiliation{QuTech, Delft University of Technology, Lorentzweg 1, 2628 CJ Delft, The Netherlands}
\author{L.~M.~K.~Vandersypen}
\affiliation{QuTech, Delft University of Technology, Lorentzweg 1, 2628 CJ Delft, The Netherlands}
\affiliation{Kavli Institute of Nanosicence, Delft University of Technology, 2600 GA Delft, The Netherlands}

\date{\today}

\begin{abstract}
We report the first complete characterization of single-qubit and two-qubit gate fidelities in silicon-based spin qubits, including cross-talk and error correlations between the two qubits. To do so, we use a combination of standard randomized benchmarking and a recently introduced method called character randomized benchmarking, which allows for more reliable estimates of the two-qubit fidelity in this system. Interestingly, with character randomized benchmarking, the two-qubit CPhase gate fidelity can be obtained by studying the additional decay induced by interleaving the CPhase gate in a reference sequence of single-qubit gates only. This work sets the stage for further improvements in all the relevant gate fidelities in silicon spin qubits beyond the error threshold for fault-tolerant quantum computation. 
\end{abstract}

\maketitle


\section*{INTRODUCTION}

With steady progress towards practical quantum computers, it becomes increasingly important to efficiently characterize the relevant quantum gates. Quantum process tomography \cite{chuang1997prescription, o2004quantum, merkel2013self} provides a way to reconstruct a complete mathematical description of any quantum process, but has several drawbacks. The resources required increase exponentially with qubit number and the procedure cannot distinguish pure gate errors from state preparation and measurement (SPAM) errors, making it difficult to reliably extract small gate error rates. Randomized benchmarking (RB) was introduced as a convenient alternative~\cite{emerson2007symmetrized, knill2008randomized, gaebler2012randomized, chow2009randomized}. It estimates the gate fidelity as a concise and relevant metric, requires fewer resources, is more robust against SPAM errors and works well even for low gate error rates.

Various randomized benchmarking methods have been investigated to extract fidelities and errors in different scenarios. 
In standard randomized benchmarking, sequences of increasing numbers of random Clifford operations are applied to one or more qubits~\cite{knill2008randomized, gaebler2012randomized}. Then, loosely speaking, the average Clifford gate fidelity is extracted from how rapidly the final state diverges from the ideally expected state as a function of the number of random Clifford operations. In interleaved randomized benchmarking, the fidelity of a particular quantum gate is obtained by interleaving that gate in a reference sequence of random Clifford gates and studying how much faster the final state deviates from the ideal case~\cite{magesan2012efficient}. Simultaneous randomized benchmarking uses simultaneously applied random Clifford operations to different qubits to characterize the degree of cross-talk~\cite{gambetta2012characterization}.

A major drawback of these traditional randomized benchmarking methods is that the number of native gates that needs to be executed in sequence to implement a Clifford operation, can rapidly increase with the qubit number. For example, it takes on average 1.5 controlled-phase (CPhase) gate and 8.25 single-qubit gates to implement a two-qubit Clifford gates \cite{corcoles2013process}. This in turns puts higher demands on the coherence time, which is still a challenge for near-term devices, and leads to rather loose bounds on the gate fidelity inferred from interleaved randomized benchmarking~\cite{magesan2012efficient, dugas2016efficiently}. Therefore, in early work characterizing two-qubit gate fidelities for superconducting qubits, the effect of the two-qubit gate projected in single-qubit space was reported instead of the actual two-qubit gate fidelity~\cite{chen2014qubit, casparis2016gatemon}. For semiconductor spin qubits, even though two-qubit Bell states have been prepared~\cite{shulman2012demonstration, watson2018programmable, zajac2018resonantly, huang2018fidelity} and simple quantum algorithms were implemented on two silicon spin qubits \cite{watson2018programmable}, the implementation issues of conventional randomized benchmarking have long stood in the way of quantifying the two-qubit gate fidelity. These limitations can be overcome either by using different native gates~\cite{huang2018fidelity} or by using a new method called character randomized benchmarking (CRB)~\cite{helsen2018new}, which allows to extract a two-qubit gate fidelity  by interleaving the two-qubit gate in a reference sequence consisting of a small number of single-qubit gates only. As an additional benefit, CRB provides detailed information on separate decay channels and error correlations.

Here we supplement standard randomized benchmarking with character randomized benchmarking for a comprehensive study of all the relevant gate fidelities of two electron spin qubits in silicon quantum dots, including the single-qubit and two-qubit gate fidelity as well as the effect of cross-talk and correlated errors on single-qubit gate fidelities. This work is of strong interest since silicon spin qubits are highly scalable, owing to their compact size ($<$ 100 nm pitch), coherence times up to tens of milliseconds and ability to leverage existing semiconductor technology~\cite{zwanenburg2013silicon, vandersypen2017interfacing}.

\section*{DEVICE AND QUBIT OPERATION}

\begin{figure}[t] 
\center{\includegraphics[width=1.0\linewidth]{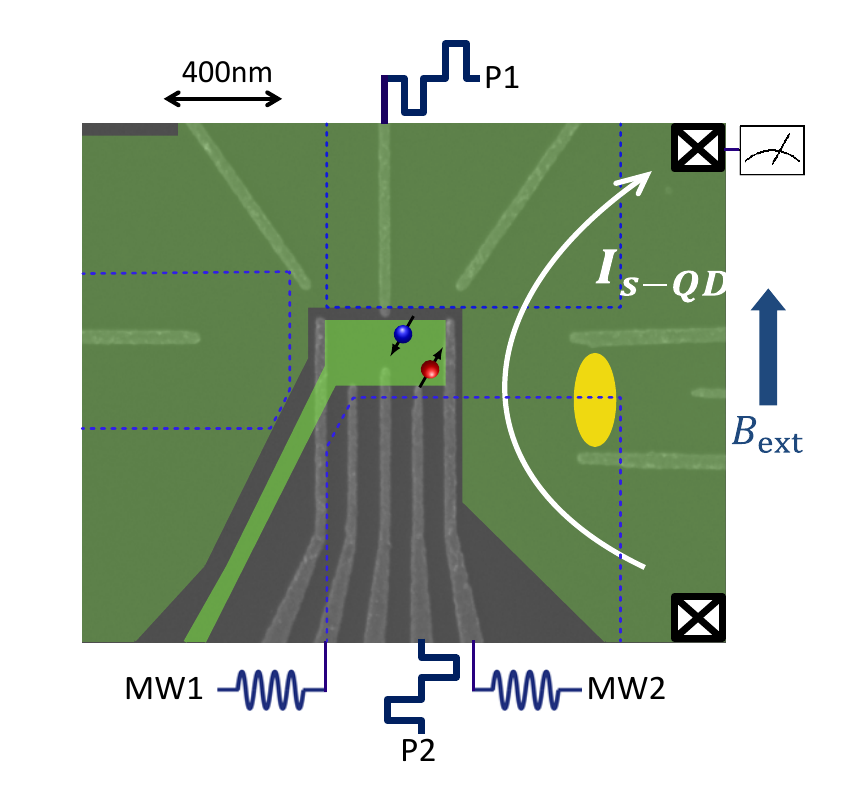}}
\caption{Device Schematic. A double quantum dot is formed in the Si/SiGe quantum well, where two spin qubits Q1 (blue spin) and Q2 (red spin) are defined. The green shaded-areas show the location of two accumulation gates on top of the double dot and the reservoir. The blue dashed lines indicate the positions of three Co micro-magnets, which form a magnetic field gradient along the qubit region. MW1 and MW2 are connected to two vector microwave sources to perform EDSR for single-qubit gates. The yellow ellipse shows the position of a larger quantum dot which is used as a charge sensor for single-shot readout. Plunger gates P1 and P2 are used to pulse to different positions in the charge stability diagram as needed for initialization, manipulation, and readout, as well as for pulsing the detuning for controlling the two-qubit gate.}
\label{fig:device}
\end{figure}

Fig.~\ref{fig:device} shows a schematic of the device, a double quantum dot defined electrostatically in a 12 nm thick Si/SiGe quantum well, 37 nm below the semiconductor surface. The device is cooled to $\sim 20$ mK in a dilution refrigerator. By applying positive voltages on the accumulation gate, a two-dimensional electron gas is formed in the quantum well. Negative voltages are applied to the depletion gates in such a way that two single electrons are confined in a double well potential~\cite{watson2018programmable}. A 617 mT magnetic field is applied in the plane of the quantum well. Two qubits, Q1 and Q2, are encoded in the Zeeman split state of the two electrons.

Single-qubit rotations rely on electric dipole spin resonance (EDSR), making use of artificial spin-orbit coupling induced by the transverse magnetic field gradient from three cobalt micro magnets fabricated on top of the gate stack~\cite{pioro2008electrically}. The longitudinal magnetic field gradient leads to well-separated spin resonance frequencies of 18.34 GHz and 19.72 GHz for Q1 and Q2 respectively. The rotation axis in the $\hat{x}-\hat{y}$ plane is set by the phase of the on-resonance microwave drive, while rotations around the $\hat{z}$ axis are implemented by changing the rotating reference frame in software~\cite{vandersypen2005nmr}.

We use the CPhase gate as the native two-qubit gate. An exchange interaction $J(\varepsilon)$ is switched on by pulsing the detuning $\varepsilon$ (electrochemical potential difference) between the two quantum dots, such that the respective electron wave functions overlap. Due to the large difference in qubit energy splittings, the flip-flop terms in the exchange Hamiltonian are ineffective and an Ising interaction remains~\cite{meunier2011efficient, veldhorst2015two, watson2018programmable, zajac2018resonantly}. The resulting time evolution operator in the standard $\{\ket{00}, \ket{01}, \ket{10}, \ket{11}\}$ basis is given by
\begin{equation}
U_{J}(t)=
\begin{pmatrix*}
	1 & 0 & 0 & 0\\
	0 & \phantom{-}e^{iJ(\epsilon)t/2\hbar} & 0 & 0\\
	0 & 0 & \phantom{-}e^{iJ(\epsilon)t/2\hbar} & 0\\
	0 & 0 & 0 & 1
\end{pmatrix*}.
\end{equation}
Choosing $t=\pi\hbar/J(\epsilon)$ and adding single-qubit $\hat{z}$ rotations on both qubits, we obtain a CPhase operator
\begin{equation}
Z_1\left(-\frac{\pi}{2}\right) Z_2(-\frac{\pi}{2}) U_{J}\!\left(\frac{\pi\hbar}{J(\epsilon)}\right) 
\!=\!
\begin{pmatrix*}[r]
	1 & 0 & 0 & 0\\
	0 & \phantom{-}1 & 0& 0\\
	0 & 0 & \phantom{-}1 & 0\\
	0 & 0 & 0 & -1
\end{pmatrix*}\!,
\end{equation}
with $Z_i(\theta)$ a $\hat{z}$ rotation of qubit $i$ over an angle $\theta$.

Spin initialization and single-shot readout of Q2 are realized by energy-selective tunnelling~\cite{elzerman2004single}. Q1 is initialized to its ground spin state by fast spin relaxation at a hotspot~\cite{srinivasa2013simultaneous}. For read-out, the state of Q1 is mapped onto Q2 using a conditional $\pi$ rotation~\cite{veldhorst2015two, watson2018programmable}, which enables extracting the state of Q1 by measuring Q2. Further details on the measurement setup are provided in Appendix~\ref{app:setup}.

\section*{INDIVIDUAL AND SIMULTANEOUS RANDOMIZED BENCHMARKING}
 
 \begin{figure*}[t] 
\center{\includegraphics[width=1.0\linewidth]{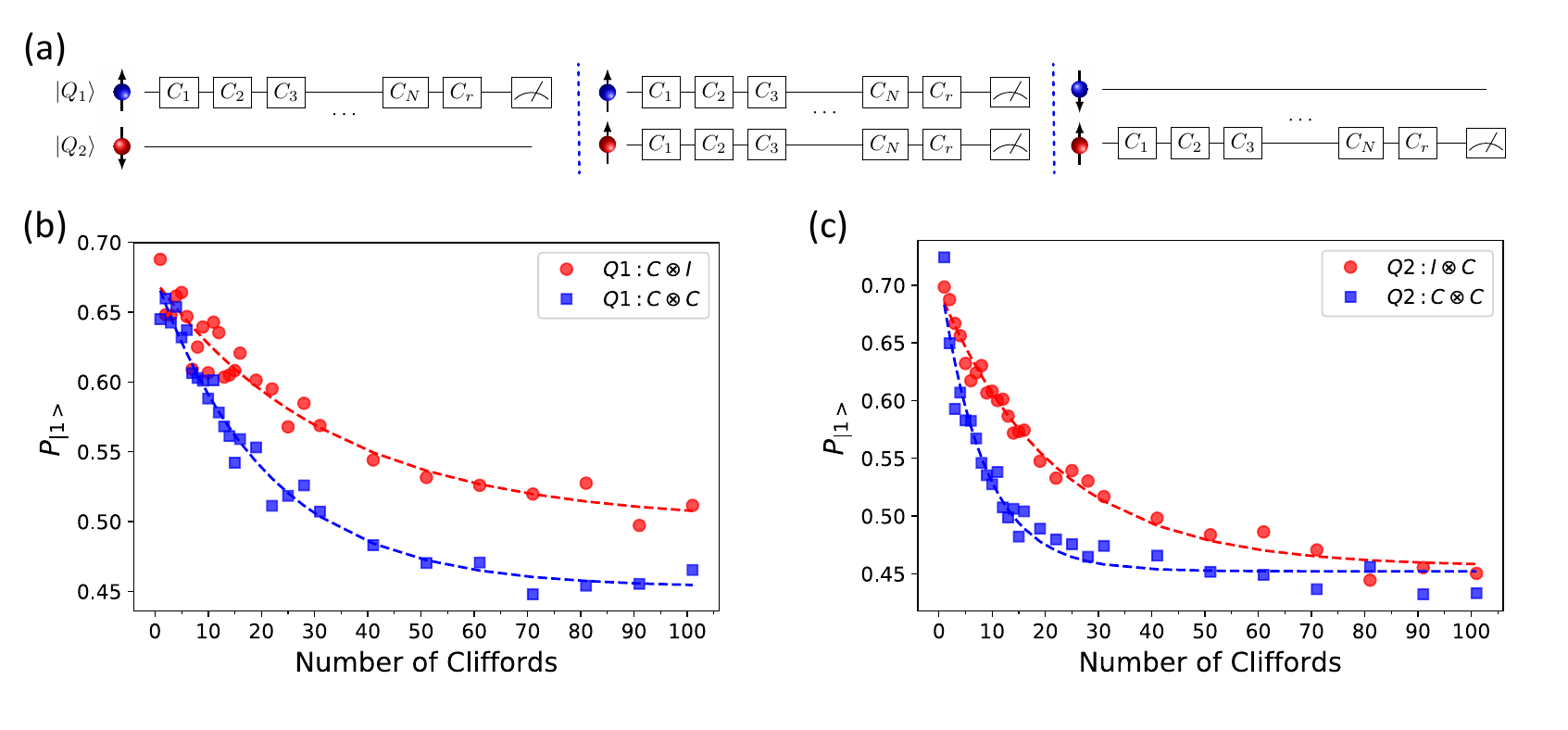}}
\caption{Individual and simultaneous standard randomized benchmarking. (a) Circuit diagrams for individual single-qubit RB on Q1 (left) and Q2 (right), and simultaneous single-qubit RB (middle). (b) Probability for obtaining outcome 0 upon measurement in the $\sigma_z\otimes{I}$ basis as a function of the number of single-qubit Clifford operations. For the red circles, Q2 is idle while a Clifford operation is applied to Q1 ($C\otimes{I}$). For the blue squares, random Clifford operations are applied to Q1 when Q2 simultaneously ($C\otimes{C}$). For each data point, we sample 32 different random sequences, which are each repeated 100 times. Dashed lines are fit to the data with a single exponential. A constant offset of -0.06 is added to the blue curve in order to compensate for a change in read-out fidelities between the two data sets, making comparison of the two traces easier. Without SPAM errors, the datapoints would decay from 1 to 0.5. (c) Analogous single-qubit RB data for Q2, with Q1 idle (red circles) and subject to random Clifford operations (blue squares). A constant offset of -0.05 is added to the blue datapoints. Throughout, single-qubit Clifford operations are generated by the native gate set $\{I, X(\pi), Y(\pm\pi), X(\pm\pi/2), Y(\pm\pi/2)\}$.}
\label{fig:simRB}
\end{figure*}

In standard randomized benchmarking, sequences of random Clifford operations are applied to a number of target qubits, followed by a final Clifford operation that, in the absence of errors, maps the qubits' state back to the initial state. Twirling one or more qubits via random Clifford operations symmetrizes the effects of noise such that  the qubits are effectively subject to a depolarizing channel. The probability $P$ that the qubits returns to the initial state then decays exponentially with the number of Clifford operations $m$, under broad assumptions~\cite{magesan2011scalable, magesan2012characterizing, wallman2018randomized}. By fitting the decay curve to 
\begin{equation}
P= A{\alpha^m+B},
\label{eq:RBdecay}
\end{equation}
where only $A$ and $B$ depend on the state preparation and measurement, the average fidelity of a Clifford operation can be extracted in terms of the depolarizing parameter $\alpha$ as
\begin{equation}
F_{avg}=1-(1-\alpha)\frac{d-1}{d},
\label{eq:Favg}
\end{equation}
where $d=2^{N}$ and $N$ is the number of qubits.\\

In the present two-qubit system, we first perform standard RB on each individual qubit (red data points in Fig.~\ref{fig:simRB}), finding $F_{avg} = 98.50\pm0.05\%$ for Q1 and $F_{avg} = 97.72\pm0.03\%$ for Q2 (all uncertainties are standard deviations). By dividing the error rate over the average number of single-qubit gates needed for a Clifford operation, we extract average single-qubit gate fidelities of $99.20\pm0.03\%$ for Q1 and $98.79\pm0.02\%$ for Q2. 

In order to assess the effects of crosstalk, we next perform single-qubit RB while simultaneously applying random Clifford operations to the other qubit (Fig.~\ref{fig:simRB} blue data points). Following~\cite{gambetta2012characterization}, we denote the corresponding depolarizing parameter for qubit $i$ while twirling qubit $j$ as $\alpha_{i|j}$. In contrast to standard RB which is insensitive to SPAM errors, we have to assume here that operations on one qubit do not affect the read-out fidelity of the other qubit~\cite{gambetta2012characterization}. Comparing with individual single-qubit randomized benchmarking results, we find that simultaneous RB decreases the average Clifford fidelity for Q1 by 0.8\% to $97.67\pm0.04\%$ while the fidelity for Q2 decreases by 3.5\% to $94.26\pm0.10\%$. Since the difference in qubit frequencies of 1.38 GHz is almost three orders of magnitude larger than the Rabi frequencies ($\sim$ 2 MHz), this crosstalk is not due to limited addressability. Furthermore, the cross-talk on Q2 persists when the drive on Q1 is applied off-resonantly, hence it is an effect of the excitation and not a result of twirling Q1. Attempting to understand how the excitation leads to undesired cross-talk, we performed detailed additional studies (see~\cite{watson2018programmable} and Appendix~\ref{app:crosstalk}), ruling out a number of other possible sources of cross-talk, including the AC Stark effect, heating and residual coupling between the qubits. Finally, cross-talk in the experimental setup is likely to be symmetric, so the observed asymmetry indicates that the microscopic details of the quantum dots must also play a role.

\section*{TWO-QUBIT RANDOMIZED BENCHMARKING}

\begin{figure}[t] 
\center{\includegraphics[width=1.0\linewidth]{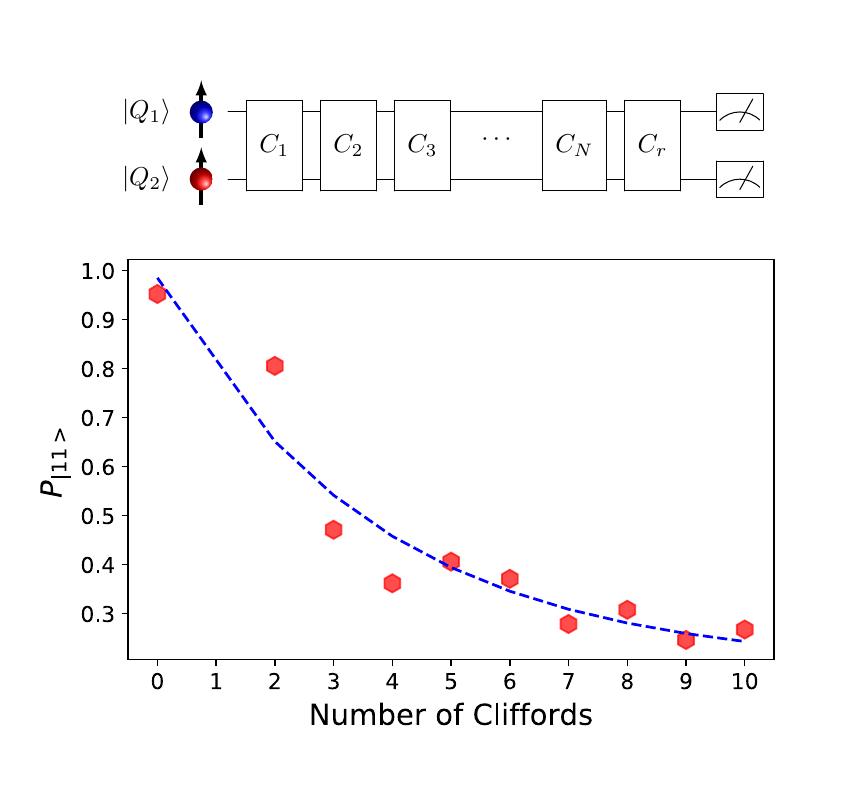}}
\caption{Two-qubit Clifford Randomized Benchmarking. Probability for obtaining outcome 11 upon measurement in the $\sigma_z\otimes{\sigma_z}$ basis, starting from the initial state $\ket{11}$, as a function of the number of two-qubit Clifford operations. As the native gate set, we use $\{I, X(\pi), Y(\pm\pi), X(\pm\pi/2), Y(\pm\pi/2),\mbox{CPhase}\}$. The elements of the two-qubit Clifford group fall in four classes of operations, the parallel single-qubit Clifford class, the CNOT-like class, the iSWAP-like class and the SWAP-like class. They are compiled by single-qubit gates plus 0, 1, 2 and 3 CPhase gates respectively. For each data point, we sample 30 random sequences, which are each repeated 100 times. The dashed line is a fit to the data with a single exponential.}
\label{fig:twoqubitRB}
\end{figure}

To characterize two-qubit gate fidelities, the Clifford group is expanded to a four-dimensional Hilbert space. We first implement standard two-qubit RB, sampling Clifford operations from the 11520 elements in the two-qubit Clifford group. Each two-qubit Clifford operation is compiled from single-qubit rotations and the two-qubit CPhase gate, requiring on average 8.25 single-qubit rotations around $\hat{x}$ or $\hat{y}$ and 1.5 CPhase gate. The measured probability to successfully recover the initial state is shown in Fig.~\ref{fig:twoqubitRB}. From a fit to the data using Eq.~\ref{eq:RBdecay} and applying Eq.~\ref{eq:Favg}, we extract an average two-qubit Clifford fidelity $F_{avg}$ of $82.10 \pm 2.75\%$. 

The large number of native gates needed to implement a single two-qubit Clifford gate, leads to a fast saturation of the decay, within about eight Clifford operations, leading to a large uncertainty on the two-qubit Clifford fidelity estimate. In addition, this fast saturation makes it difficult to assess whether gate-dependent errors are present~\cite{wallman2018randomized, carignan2018randomized, proctor2017randomized}. Importantly, interleaving a specific gate in a fast decaying reference sequence also yields a rather unreliable estimate of the interleaved gate fidelity. In the present case, interleaving a CPhase gate in the reference sequence of two-qubit Clifford operations is not a viable strategy to extract the CPhase gate fidelity. Furthermore, the compilation of Clifford gates into two different types of native gates -- single-qubit gates and the CPhase gate -- makes it impossible to confidently extract the fidelity of any of the native gates, such as the CPhase gate, by itself. This is different from a recent experiment on silicon spin qubits where only a single physical native gate was used, the conditional rotation, in which case the error per Clifford operation can be divided by the average number of conditional rotations per Clifford operation for estimating the error per conditional rotation~\cite{huang2018fidelity}. 

As a first step to obtain quantitative information on the CPhase gate fidelity, we implement a simplified version of interleaved RB, which provides the fidelities of the two-qubit gate projected in various single-qubit subspaces, as was done earlier for superconducting transmon qubits~\cite{chen2014qubit} and hybrid gatemon qubits~\cite{casparis2016gatemon}. In this protocol, the CPhase gate is interleaved in a reference sequence of single-qubit Clifford operations. When applying a CPhase gate, we can (arbitrarily) consider one qubit the control qubit and the other the target qubit. When the control qubit is $\ket{1}$, the target qubit ideally undergoes a $\pi$ rotation around the $\hat{z}$ axis. With the control in $\ket{0}$, the target qubit ideally remains fixed (Identity operation). Therefore, projected in the subspace corresponding to the target qubit, this protocol interleaves either a $Z(\pi)$ rotation or the identity operation in a single-qubit RB reference sequence applied to the target qubit.
The decay of the return probability for interleaved RB is also expected to follow Eq.~\ref{eq:RBdecay}. 
The fidelity of the interleaved gate is then found from the depolarizing parameter $\alpha$ for the interleaved and reference sequence, as
\begin{equation}
F_{gate}=1-\left(1-\frac{\alpha_{interleaved}}{\alpha_{reference}}\right) \frac{d-1}{d}.
\label{eq:interleavedfidelity}
\end{equation}
From the experimental data, we find CPhase fidelities projected in single-qubit space of 91\% to 95\%, depending on which qubit acts as the control qubit for the CPhase, and which eigenstate it is in (see Appendix~\ref{app:projected}).

\section*{CHARACTER RANDOMIZED BENCHMARKING}

\begin{figure*}[t]
\center{
\includegraphics[width=1.0\linewidth]{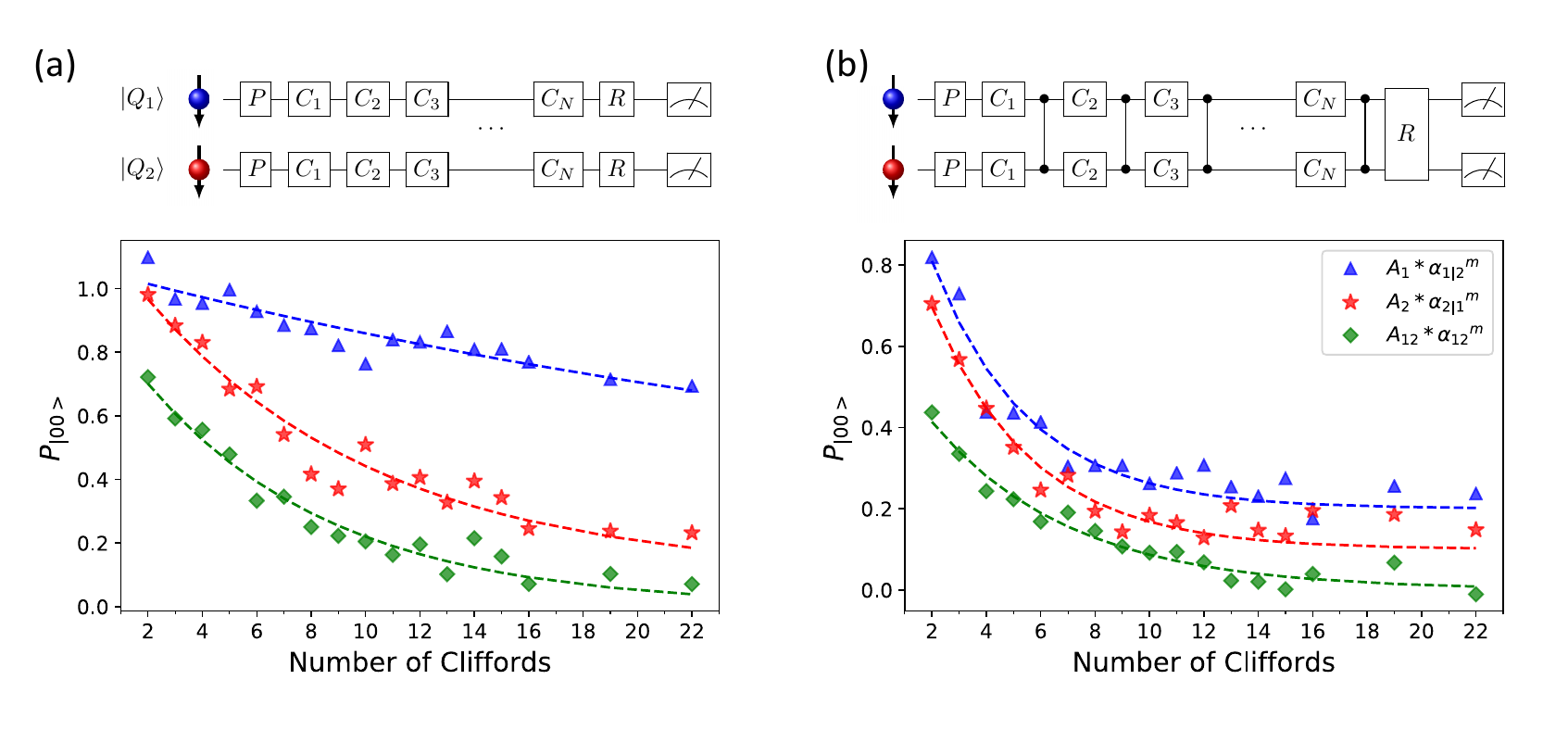}}
\caption{Character randomized benchmarking. (a) Reference CRB experiment. 
The probabilities $P_1$ (blue triangles), $P_2$ (red stars) and $P_3$ (green diamonds), obtained starting from the initial state $\ket{00}$ followed by a Pauli operation, as a function of the number of subsequent single-qubit Clifford operations simultaneously applied to both qubits (see the schematic of the pulse sequence). As the native gate set, we use $\{I, X(\pi), Z(\pm\pi), X(\pm\pi/2), Z(\pm\pi/2),\mbox{CPhase}\}$.  For each of the 16 Pauli operators, we apply 40 different  random sequences, each with 20 repetitions. The dashed lines are fits to the data with a single exponential. Without SPAM errors, the datapoints would decay from 1 to 0. (b) Interleaved CRB experiment. This experiment is performed in an analogous way to the reference CRB experiment, but with a two-qubit CPhase gate interleaved after each Clifford pair, as seen in the schematic of the pulse sequence. The traces are offset by an increment of 0.1 for clarity.
}
\label{fig:CRB}
\end{figure*}

In order to properly characterize the two-qubit CPhase fidelity, we experimentally demonstrate a new approach to RB called character randomized benchmarking (CRB)~\cite{helsen2018new}. CRB is a powerful generic method that extends randomized benchmarking in a rigorous manner, making it possible to extract average fidelities from groups beyond the multi-qubit Clifford group while keeping the advantages of standard RB such as resistance to SPAM errors. The generality of CRB allows one to start from (a subset of) the natives gates of a particular device and then design an RB experiment tailored to that set. This can strongly reduce compilation overhead and gate dependent noise, a known nuisance factor in standard RB~\cite{wallman2018randomized, carignan2018randomized, proctor2017randomized}. Moreover, since the accuracy of interleaved randomized benchmarking depends on the fidelity of the reference gates~\cite{magesan2012efficient, dugas2016efficiently}, performing (through CRB) interleaved RB with a reference group generated by high fidelity gates can significantly improve the utility of interleaved RB.

Character randomized benchmarking requires us to average over two groups (the second one usually being a subgroup of the first). The first group is the ``benchmark group". It is for the gates in this group that CRB yields the average fidelity. The second group is the ``character group". CRB works by performing standard randomized benchmarking using the benchmark group but augments this by adding a random gate from the character group before each RB gate sequence. By averaging over this extra random gate, but weighting the average by a special function known from representation theory as a character function, it guarantees that the average over random sequences can always be fitted to a single exponential decay, even when the benchmark group is not the multi-qubit Clifford group and even in the presence of SPAM errors.

Guided by the need for high reference fidelities, we choose for our implementation of CRB the benchmark group to be the parallel single-qubit Clifford group ($C\otimes C$, the same as in standard simultaneous single-qubit RB) and the two-qubit Pauli group as the character group (see~\cite{helsen2018new} for more information on why this is a good choice for the character group). It is non-trivial that the $C\otimes C$ group allows us to get information on two-qubit gates, since parallel single-qubit Clifford operations cannot fully depolarize the noise in the full two-qubit Hilbert space.
In fact, for simultaneous single-qubit RB there are three depolarizing channels, each acting in a different subspace of the Hilbert space of density matrices, spanned by $I\otimes\sigma_i$, $\sigma_i\otimes{I}$, and $\sigma_i\otimes\sigma_i$, with $I$ the identity operator and $\sigma_i$ one of the Pauli operators. The three decay channels are reflected in the recovery probability for the final state, which is now described by
\begin{equation}
P_{C \otimes C}= A_1 {\alpha_{1|2}}^m+A_2 {\alpha_{2|1}}^m+A_{12} {\alpha_{12}}^m+B,
\label{eq:simRBdecay}
\end{equation}
where $\alpha_{i|j}$ is again the depolarizing parameter for qubit $i$ while simultaneously applying random Clifford operations to qubit $j$, and $\alpha_{12}$ is the depolarizing parameter for the two-qubit parity ($\{\ket{00},\ket{11}\}$ versus $\{\ket{01},\ket{10}\}$). We note that if the errors acting on both qubits are uncorrelated, then $\alpha_{12} = \alpha_{1|2} \alpha_{2|1}$~\cite{gambetta2012characterization}. The question now is how to separate the three decays. Fitting the data using a sum of three exponentials will be very imprecise. Existing approaches combine the decay of specific combinations of the probabilities of obtaining $00, 01, 10$ and $11$ upon measurement, but suffer from SPAM errors \cite{gambetta2012characterization}. As discussed above, CRB offers a clean procedure for extracting the individual decay rates that is immune to SPAM errors and does not incur additional overhead.

Concretely, CRB here proceeds as follows: (1) the two-qubit system is initialized to $\ket{00}$, then (2) one random Pauli operator on each qubit is applied to prepare the system in a state $\ket{\phi_1\phi_2}$ (one of $\ket{00}$, $\ket{01}$, $\ket{10}$, and $\ket{11}$), followed by (3) a random sequence of simultaneously applied single-qubit Clifford operators. In practice, the random Pauli operator is absorbed in the first Clifford operation, making the Pauli gates effectively noise-free. A final Clifford operation is applied which ideally returns the system to the state $\ket{\phi_1\phi_2}$ and finally (4) both qubits are measured. Each random sequence is repeated to collect statistics on the probability $P_{\phi_1\phi_2}$ of obtaining measurement outcome $00$ when starting from $\ket{\phi_1\phi_2}$ (note that each $P_{\phi_1\phi_2}$ averages over 4 Pauli operations). We combine these probabilities according to their character (see Appendix~\ref{app:math} for more details) to obtain three fitting parameters,
\begin{equation}
\begin{split}
P_1=P_{00}-P_{01}+P_{10}-P_{11},\\
P_2=P_{00}+P_{01}-P_{10}-P_{11},\\
P_3=P_{00}-P_{01}-P_{10}+P_{11}.
\end{split}
\end{equation}
Each of these three fitting parameters is expected to decay as a single exponential, isolating one of the decay channels in Eq.~\ref{eq:simRBdecay}:
\begin{equation}
\begin{split}
P_1=A_{1} {\alpha_{1|2}}^m,\\
P_2=A_{2} {\alpha_{2|1}}^m,\\
P_3=A_{12} {\alpha_{12}}^m.
\end{split}
\end{equation}
Note that there is no constant offset $B$. This is also a feature of CRB. The three experimentally measured probabilities are shown in Fig.~\ref{fig:CRB}a. These contain a lot of useful information, including not only the separate depolarizing parameters but also the averaged CRB reference fidelity and information on error correlations. The blue (red) curve shows the decay in the subspace corresponding to Q1 (Q2), spanned by $\sigma_i\otimes{I}$ ($I\otimes\sigma_i$). The green curve shows the decay in the subspace spanned by $\sigma_i\otimes\sigma_j$. This decay can be interpreted as the parity decay. The fitted depolarizing parameters are $\alpha_{1|2} = 0.9738 \pm 0.0008, 
\alpha_{2|1} = 0.8902 \pm 0.0020$ and $\alpha_{12} = 0.8652 \pm 0.0022$.

The average CRB depolarizing parameter can be found from the separate depolarizing parameters as
\begin{equation}
P= \frac{3}{15} {\alpha_{1|2}} + \frac{3}{15} {\alpha_{2|1}} + \frac{9}{15} {\alpha_{12}},
\end{equation}
where the weights are proportional to the dimension of the corresponding subspaces of the 16-dimensional Hilbert space of two-qubit density matrices. We obtain a reference CRB fidelity of $91.9 \pm 0.1 \%$, which represents the fidelity of two simultaneous single-qubit Clifford operators ($C\otimes{C}$).

Finally, from the three depolarizing parameters in Eq.~\ref{eq:simRBdecay}, we can infer to what extent errors occur independently on each qubit or exhibit correlations between the two qubits. The fact that $\alpha_{12}-\alpha_{1|2}\alpha_{2|1}= -0.0017 \pm 0.0031$ indicates that the errors are essentially independent. 

Next we perform the interleaved version of CRB, for which we insert a CPhase gate after each single-qubit Clifford pair. Fig.~\ref{fig:CRB}b shows the three corresponding experimentally measured decays. The fitting parameters we extract now reflect the combined errors from a single-qubit Clifford pair followed by a CPhase gate. The fitted depolarizing parameters are $\alpha_{1|2} = 0.7522 \pm 0.0060, \alpha_{2|1} = 0.7623 \pm 0.0053$, and $\alpha_{12} = 0.8226 \pm 0.0030$. As can be expected, the three decays lie closer together than those for reference CRB: not only does the additional CPhase gate contribute directly to all three decays, it also mixes the three subspaces. From the depolarizing parameters in interleaved and reference CRB measurement, we use Eq.~\ref{eq:interleavedfidelity} to isolate the fidelity of the CPhase gate, now in two-qubit space as desired, yielding $92.0\pm0.5 \%$.

The dominant errors in the CPhase gate arise from nuclear spin noise and charge noise. In natural silicon, the abundance of Si$^{29}$ atoms is about 4.7\%, and the Si$^{29}$ nuclear spins dephase the electron spin states due to the hyperfine interaction~\cite{zwanenburg2013silicon}. Charge noise modulates the overlap of the two electron wave functions, and thus also the  two-qubit coupling strength. In the present device, we could not access the symmetry point where the coupling strength is to first order insensitive to the detuning of the double dot potential~\cite{martins2016noise, reed2016reduced}, hence charge noise directly (to first order) affects the two-qubit coupling strength. 

\section*{CONCLUSIONS}

Character randomized benchmarking provides a new method to effectively characterize multi-qubit behaviour. It combines the advantages of simultaneous randomized benchmarking and interleaved randomized benchmarking, and gives tighter bounds on the fidelity number than standard interleaved randomized benchmarking due to its simpler compilation. CRB is useful in a wide variety of settings, far beyond the particular case studied here. The general approach to exploiting CRB is to start from a set of native gates that can be implemented easily and with high fidelity, and to construct a suitable reference sequence based on this set.  The decay for the reference sequence contains any number of exponentials, which can be separated without suffering from SPAM errors and which provide relevant additional information, in the present case on the fidelity of simultaneously applied gates, cross-talk and on noise correlations. Comparison with interleaved CRB allows one to extract the fidelity of specific gates of interest.
 
We perform the first comprehensive study of the single-qubit, simultaneous single-qubit and two-qubit gate fidelities for semiconductor qubits, where the use of CRB, which allows for a compact reference sequence, was essential for extracting a reliable two-qubit gate fidelity. Summarizing, independent single-qubit gate fidelities are around $99\%$ in this system, these drop to $98.8\%$ for qubit 1 and to $96.9\%$ for qubit 2 when simultaneously twirling the other qubit, and the two-qubit CPhase fidelity is around 91\%. We expect that by working in an isotopically purified Si$^{28}$/SiGe substrate and performing the two-qubit gate at the symmetry point, a CPhase gate fidelity above the fault-tolerant threshold ($>99\%$) can be reached. A recent report on the fidelity of controlled rotations in Si/SiO$_2$ quantum dots already comes close to this threshold~\cite{huang2018fidelity}. With further improvements in charge noise levels, two-qubit gate fidelities above $99.9\%$ are in reach.\\

\section*{ACKNOWLEDGMENTS}
This research was sponsored by the Army Research Office (ARO) under grant numbers W911NF-17-1-0274 and W911NF-12-1-0607. The views and conclusions contained in this document are those of the authors and should not be interpreted as representing the official policies, either expressed or implied, of the ARO or the US Government. The US Government is authorized to reproduce and distribute reprints for government purposes notwithstanding any copyright notation herein. Development and maintenance of the growth facilities used for fabricating samples is supported by DOE (DE- FG02-03ER46028). We acknowledge the use of facilities supported by NSF through the University of Wisconsin-Madison MRSEC (DMR-1121288). J.H. and S.W. are supported by an NWO VIDI Grant (SW), an ERC Starting Grant (SW), and the NWO Zwaartekracht QSC. We acknowledge useful discussions with the members of the Vandersypen group, and technical assistance by R. Schouten and R. Vermeulen.

\appendix

\section{MEASUREMENT SETUP}\label{app:setup}
\begin{table*}
\begin{tabular}{|c|c|c|c|c|c|c|c|c|c|c|c|c|c|c|c|c|}
\hline
$\sigma\backslash P$ & $II$ & $\sigma_zI$ & $I \sigma_z$ & $\sigma_z\sigma_z$ & $\sigma_xI$ & $I \sigma_x$ & $\sigma_x\sigma_x$ & $\sigma_yI $ & $I \sigma_y$ & $\sigma_y\sigma_y$ & $\sigma_z\sigma_x$ & $\sigma_x\sigma_z$ & $\sigma_z\sigma_y$ & $\sigma_y\sigma_z$ & $\sigma_x\sigma_y$ & $\sigma_y\sigma_x$\\
\hline\hline
$\sigma_zI$ &  $1$ & $1$ & $1$ & $1$ & $-1$ & $1$ & $-1$ & $-1$ & $1$ & $-1$ & $1$ & $-1$ & $1$ & $-1$ & $-1$ & $-1$\\
\hline
$I\sigma_z$ &  $1$ & $1$ & $1$ & $1$ & $1$ & $-1$ & $-1$ & $1$ & $-1$ & $-1$ & $-1$ & $1$ & $-1$ & $1$ & $-1$ & $-1$\\
\hline
$\sigma_z\sigma_z$ &  $1$ & $1$ & $1$ & $1$ & $-1$ & $-1$ & $1$ & $-1$ & $-1$ & $1$ & $-1$ & $-1$ & $-1$ & $-1$ & $1$ & $1$\\
\hline
\end{tabular}
\caption{Values for the character function $\chi_{P}(\sigma)$ for $P \in \{(\sigma_z \otimes I ),(I \otimes \sigma_z ),(\sigma_z \otimes \sigma_z )\}$.}\label{box:char}
\end{table*}
The measurement setup is the same as the one used in~\cite{watson2018programmable}. We summarize here a few key points.
The gates P1 and P4 are connected to arbitrary waveform generators (AWG, Tektronix 5014C) via coaxial cables. Applying DC voltage pulses to these two gates moves the system through different positions in the charge stability diagram for initialization, operation and read-out. Voltage pulses applied to these two gates are used to pulse the detuning between the two quantum dots, thereby turning on and off the controlled-phase gate. Gates P2 and P3 are connected to vector microwave sources (Keysight E8267D) for achieving EDSR. Each microwave source has two I/Q input channels, connected to two channels on the master AWG, which controls the clock of the entire system and triggers all the other instruments. The frequency, phase and duration of the microwave bursts are thus controlled by I/Q modulation. In addition, we use pulse modulation to obtain a larger on/off ratio of the microwave bursts than is possible using I/Q modulation only. A digitizer card (Spectrum M4i.44) installed inside the measurement computer is used to record the current traces of the sensing quantum dot at a sampling rate $\sim$ 60 kHz. Each time trace is converted into a single bit value (0 or 1) by the measurement computer using threshold detection. The average over many repetitions gives us the spin-up and spin-down probabilities (0 and 1). 

\section{MATHEMATICAL BACKGROUND OF CRB}\label{app:math}

Character randomized benchmarking is a generic method for performing randomized benchmarking with finite groups other than the multi-qubit Clifford group. As mentioned in the main text, CRB requires the user to specify two finite groups: the benchmark group and the character group. In this work we chose the benchmark group to be the simultaneous single-qubit Clifford group on two qubits and the character group to be the two-qubit Pauli group. Standard RB and CRB rely on the framework of representation theory. Central to the use of RB and CRB is a powerful result called Schur's lemma. In the context of this paper Schur's lemma states that, assuming for simplicity that every gate is subject to an identical noise map $\mathcal{E}$, the average noisy RB operator $\mathcal{M}$ is of the form
\begin{equation*}
\mathcal{M}:=\!\!\!\!\!\!\sum_{\substack{G_1,\ldots G_m\\\in C\otimes C}}\!\!\! \!G_{inv}\mathcal{E}G_m\cdots \mathcal{E}G_1 \!=\!\begin{bmatrix}1 &  &  & \\  &\!\!\! \alpha_{1|2}\mathbb{I}_{1|2}\!\!\!&  &\\& &\!\!\! \alpha_{2|1}\mathbb{I}_{2|1}\! & \\&   &&\!\!\! \alpha_{12}\mathbb{I}_{12}\!\end{bmatrix}^m\!\!\!\!\!,
\end{equation*}
where we are describing all quantum channels in the Pauli Transfer Matrix picture, i.e. $\mathcal{M}_{i,j} = \text{Tr}(\sigma_i \mathcal{M}(\sigma_j))/2$ where $\sigma_i, \sigma_j$ are Pauli matrices. One can think of the matrix entry $\mathcal{M}_{i,j}$ as describing how much the noise map $\mathcal{M}$ maps the generalized Bloch sphere axis labeled $\sigma_j$ to the one labeled $\sigma_j$. The submatrices $\mathbb{I}_{1|2},\mathbb{I}_{1|2}$ and $\mathbb{I}_{12}$ of the matrix $\mathcal{M}$ are defined as the identity matrix on the sets of 2-qubit Pauli's of the form $\{\sigma_i\otimes I\}, \{I\otimes\sigma_i\}$ and $\{\sigma_i\otimes \sigma_j\}$ respectively. We would like to estimate the numbers $\alpha_{1|2}, \alpha_{2|1}$ and $\alpha_{12}$ individually in a way that does not depend on state preparation and measurement. To do this CRB adds an extra average over another group called the character group, which we choose to be the two-qubit Pauli group. This average is weighted by a so-called character function. This average over the Pauli group projects any initial state onto a single axis of the Bloch sphere. Which axis is projected on depends on the character function used for the weights. By selecting the correct Bloch sphere axes, we can single out the individual blocks of the matrix $\mathcal{M}$. In order to isolate the parameter $\alpha_{1|2}$ we choose to project onto the Bloch sphere axis associated to $\sigma_z\otimes I$. Concretely this means that the character averaged RB operator $\mathcal{M}$ becomes
\begin{equation*}
 \sum_{\sigma\in P_2}\!\chi_{\sigma_zI}(\sigma)\!\!\!\!\sum_{\substack{G_1,\ldots G_m\\\in C\otimes C}}\!\! G_{inv}\mathcal{E}G_m\cdots \mathcal{E}G_1\sigma = \mathcal{M}^m \mathcal{P}_{ZI}
 \end{equation*}
 where the function $\chi_{\sigma_zI}(\sigma)$ is given in the first row of Table \ref{box:char}  and the matrix $\mathcal{P}_{\sigma_zI}$ has all zero entries except on the diagonal entry corresponding to the Pauli $\sigma_z\otimes I$. By matrix multiplication we see that $\mathcal{M}^m \mathcal{P}_{\sigma_zI} = \alpha_{1|2}^m\mathcal{P}_{\sigma_zI}$. This means that the average measured survival probability in CRB, with input state $\rho$ and measurement operator $Q$ is of the form 
 \begin{equation*}
  \sum_{\sigma\in P_2}\! \chi_{ZI}(\sigma)\!\!\!\! \sum_{\substack{G_1,\ldots G_m\\\in C\otimes C}}\!\!\text{Tr}(Q G_{inv}\mathcal{E}G_m\cdots \mathcal{E}G_1(\sigma(\rho))) = A \alpha_{1|2}^m
  \end{equation*}
  where $A$ is a function of $Q$ and $\rho$. Similarly we can obtain estimates $\alpha_{2|1}$ and $\alpha_{12}$ by constructing projectors onto the Pauli operators $I\otimes\sigma_z$ and $\sigma_z\otimes \sigma_z$ respectively. The character functions for these projectors are given in rows 2 and 3 of Table \ref{box:char} respectively.

As noted in the main text, CRB is a generic procedure, which can be used beyond its application in this manuscript. Another notable example of where we suspect CRB can offer a benefit is when the device native gates are not single-qubit gates but rather two-qubit gates, as happens in~\cite{huang2018fidelity}. In this case compiling multi-qubit Cliffords is very cumbersome. In the theoretical RB literature benchmarking groups are discussed that are more suitable to this scenario such as the CNOT-dihedral group (for native CNOT gates)~\cite{cross2016scalable} and the real Clifford group (for native CPhase gates)~\cite{Hashagen2018realrandomized}. Both of these groups lead to benchmarking data that mixes two exponential decays but using the CRB approach these can be fitted individually in a reliable manner (in both cases the Pauli group is a good choice for character group, see the example in~\cite{helsen2018new} for more information).

\section{EXPERIMENTAL DETAILS FOR CRB}

The single-qubit Clifford group is commonly generated by the gate set $\{I, X(\pi), Y(\pm\pi), X(\pm\pi/2), Y(\pm\pi/2)\}$. In our experiment, we perform Z rotations by changing a qubit's reference frame in software~\cite{vandersypen2005nmr}, which makes Z rotations error-free. To benefit from this, we generate the single-qubit Clifford group by the gate set $\{I, X(\pi), Z(\pm\pi), X(\pm\pi/2), Z(\pm\pi/2)\}$ instead. Furthermore, we keep the Rabi frequency the same for all the X rotations, thus a $X(\pi)$ gate has twice the duration of a $X(\pi/2)$ gate. Combined with using X-Z compilation, we can keep the duration for all the 24 Clifford operations the same as shown below, thereby avoiding any unnecessary idle time which would quickly dephase the qubits.

\begin{center}
\begin{table}
\begin{tabular}{p{6em} p{7em} p{8em} } 
 \hline
Class         & X-Y generation          & X-Z generation \\ 
[0.5ex] 
 \hline
 \multirow{4}{*}{Pauli} & $I$ & $X, -X$\\ 
 
 & $X^2$ & $X^2$\\
 
 & $Y^2$ & $-Z, X^2, Z$\\
 
 & $Y,^2 X^2$ & $X, Z^2, X$\\ 
 \hline
 \multirow{8}{*}{$2\pi/3$} & $X, Y$ & $X, -Z, X, Z$\\ 
 
 & $X, -Y$ & $X, Z, X, -Z$\\
 
 & $-X, Y$ & $-X, -Z, X, Z$\\
 
 & $-X, -Y$ & $-X, Z, X, -Z$\\
 
 & $Y, X$ & $-Z, X, Z, X$\\
 
 & $Y, -X$ & $-Z, X, Z, -X$\\
 
 & $-Y, X$ & $Z, X, -Z, X$\\
 
 & $-Y, -X$ & $Z, X, -Z, -X$\\
  
  \hline
  \multirow{6}{*}{$\pi/2$} & $X$ & $-Z, X, Z, X, -Z$\\
 
 & $-X$ & $Z, -X, -Z, -X, Z$\\
 
 & $Y$ & $X, Z, -X$\\
 
 & $-Y$ & $X, -Z, -X$\\
 
 & $-X, Y, X$ & $-X, Z^2, -X, -Z$\\
 
 & $-X, -Y, X$ & $-X, -Z^2, -X, Z$\\
  
  \hline
  \multirow{6}{*}{Hadamard} & $X^2, Y$ & $X, -Z, X$\\ 
 
 & $X^2, -Y$ & $X, Z, X$\\
 
 & $Y^2, X$ & $-Z, X, Z, X, Z$\\
 
 & $Y^2, -X$ & $-Z, X, Z, -X, -Z$\\
 
 & $X, Y, X$ & $X^2, Z$\\
 
 & $-X, Y, -X$ & $-X^2, -Z$\\ [1ex] 
 \hline
\end{tabular}
\caption{Compilation of the single-qubit Clifford group with X/Y rotations and X/Z rotations. Here $(-)K$ and $(-)K^2$ denote $K(\pm\pi/2)$ and $K(\pm\pi)$ gates ($K = X, Y, Z$) respectively.}\label{box:fidelity}
\end{table}
\end{center}

\section{Comparison of standard and character interleaved two-qubit RB}\label{app: bounds}
 
Although it often goes unmentioned, the estimate for the fidelity of an interleaved gate given in Eq.~\ref{eq:interleavedfidelity} is only exact when the qubit noise is exactly depolarizing. In the presence of other types of noise (such as dephasing or calibration errors) this number gives only upper and lower bounds on the fidelity of the interleaved gate. First upper and lower bounds were given in~\cite{magesan2012efficient} and recently optimal upper and lower bounds were given in~\cite{dugas2016efficiently}. These bounds depend strongly on the fidelity of the gates in the reference sequence, in particular they scale as $O(\sqrt{1-\alpha_{\mathrm{ref}}})$ where $\alpha_{\mathrm{ref}}$ is the reference RB decay constant. This means that our implementation of CRB, which uses only single-qubit gates for the reference experiment, has a significant advantage over standard two-qubit interleaved RB also in this respect. We can illustrate this advantage by considering a hypothetical standard two-qubit interleaved experiment with interleaved CZ gate. Recall from Eq. \ref{fig:twoqubitRB} that standard two qubit RB (here considered as a reference experiment) yielded a reference fidelity of $82\%$ and thus a depolarizing parameter of $\alpha_{2,\mathrm{ref}} = 0.73$ (suppressing uncertainty for the sake of this exercise). Assuming an interleaved CPhase fidelity of $92\%$ (which is what we extracted from the CRB experiment) and assuming that the error on a hypothetical interleaved two qubit RB experiment scales multiplicatively (optimistic given the possibility of calibration errors) we estimate that a hypothetical two qubit interleaved RB experiment would have a depolarizing parameter of $\alpha_{2,\mathrm{int}}$. Using the optimal bounds calculated in~\cite{dugas2016efficiently} this would mean we can only guarantee that the fidelity of the interleaved gate lies in the range $[0.58,1]$. From the CRB experiment we can however guarantee that the fidelity of the interleaved gate lies in the range $[0.69,1]$, a significant improvement even in the absolute worst case scenario discussed in~\cite{dugas2016efficiently}. 

We would also like to note that the bounds given in~\cite{magesan2012efficient,dugas2016efficiently} significantly overestimate the range of possible interleaved gate fidelities if more is known about the noise process. If for instance the noise on the reference gates is assumed to be dominated by stochastic errors (as opposed to coherent errors due to mis-calibration) then the upper and lower bounds can be made significantly tighter. This coincides with experimental consensus that interleaved RB generally gives good estimates of the interleaved gate fidelity. However, since single qubit gates will typically suffer less from calibration errors than compiled two qubit gates we argue that interleaved CRB will yield sharper upper and lower bounds on the interleaved gate fidelity than standard interleaved RB when more is known about the noise process.

\section{Interleaved RB projected in single-qubit space}\label{app:projected}
\begin{center}
\begin{table}[h]
\begin{tabular}{p{2.5em} p{5em} p{2.5em} p{5em} p{6em} } 
 \hline
$Q_C$ & $Q_C$ state & $Q_T$  & operation & fidelity\\
[1ex] 
 \hline
 $Q_1$ & $\ket{0}$ & $Q_2$ & $I$ & $94.62\pm0.24\%$\\ 
 
 $Q_1$ & $\ket{1}$ & $Q_2$ & $Z(\pi)$ & $90.79\pm0.38\%$\\
 
 $Q_2$ & $\ket{0}$ & $Q_1$ & $I$ & $95.50\pm0.20\%$\\
 
 $Q_2$ & $\ket{1}$ & $Q_1$ & $Z(\pi)$ & $94.38\pm0.25\%$\\ [1ex] 
 \hline
\end{tabular}
\caption{CZ fidelities for different target-qubits ($Q_T$), and different states of the control-qubit ($Q_C$)}\label{box:fidelity}
\end{table}
\end{center}
\begin{figure} [h]
\center{\includegraphics[width=1.0\linewidth]{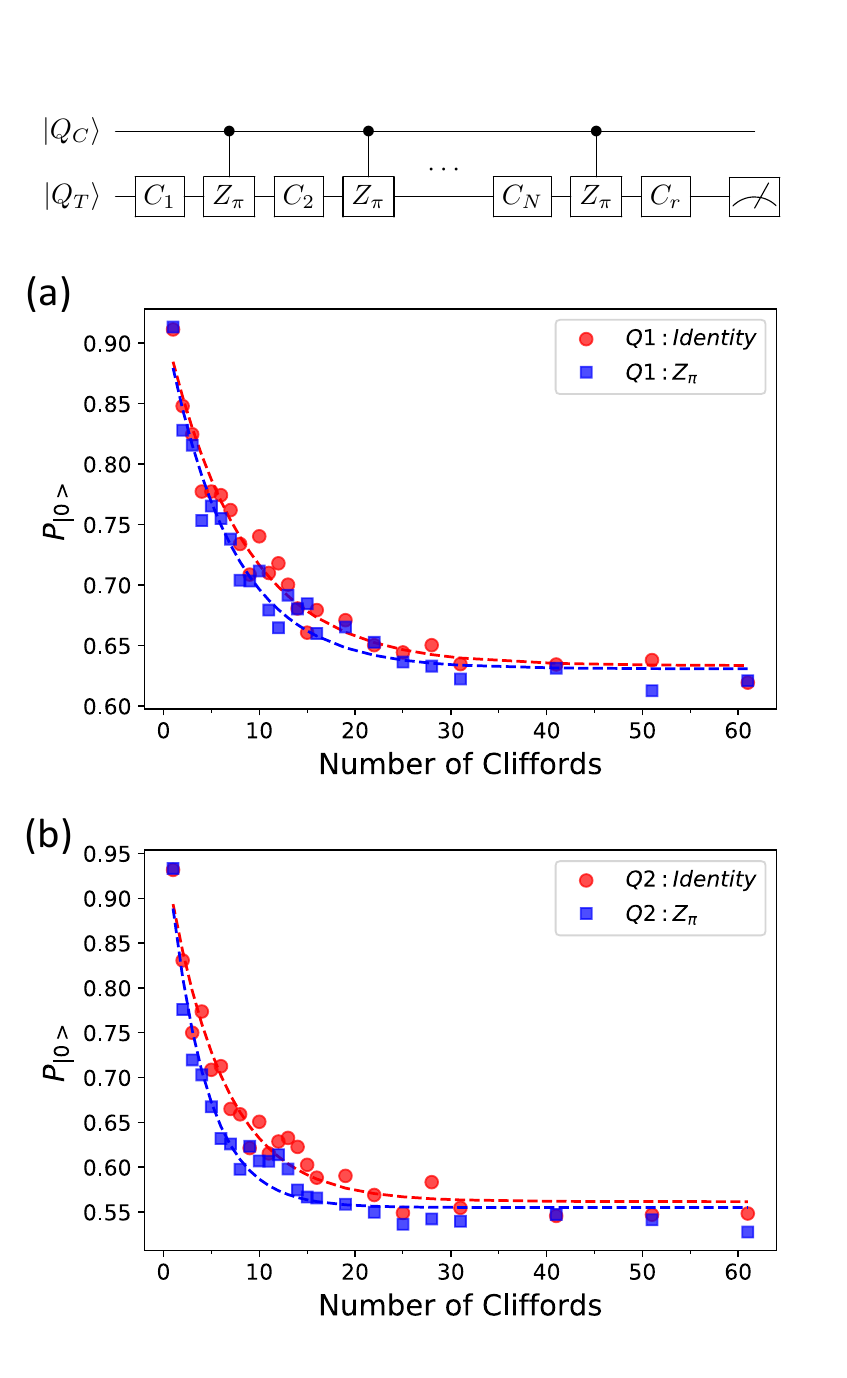}
}
\caption{Interleaved Randomized Benchmarking projected in single-qubit space. (a) Probability for obtaining outcome 0 upon measurement in the $\sigma_z\otimes{I}$ basis as a function of the number of single-qubit Clifford operations, interleaved with the CPhase operation. For the red circles (blue squares), Q2 is is in $\ket{0}$ ($\ket{1}$) so Q1 is expected to undergo the identity operation (a $Z(\pi)$ rotation). For each data point, we sample 30 different random sequences for each Clifford number, which are each repeated 100 times. Dashed lines are fits to the data with a single exponential. (b) Analogous data for Q2.}
\label{fig:projected}
\end{figure}

Fig.~\ref{fig:projected} shows experimental results for the experiment discussed in the main text where a CPhase gate is interleaved in a standard single-qubit RB sequence applied to one qubit, while the other qubit is in either $\ket{0}$ or $\ket{1}$. This experiment provides the CPhase fidelity projected in single-qubit space~\cite{chen2014qubit, casparis2016gatemon}, summarized in the table below for the four possible cases.

\section{CROSSTALK}\label{app:crosstalk}

\begin{figure} [h]
\center{\includegraphics[width=1.0\linewidth]{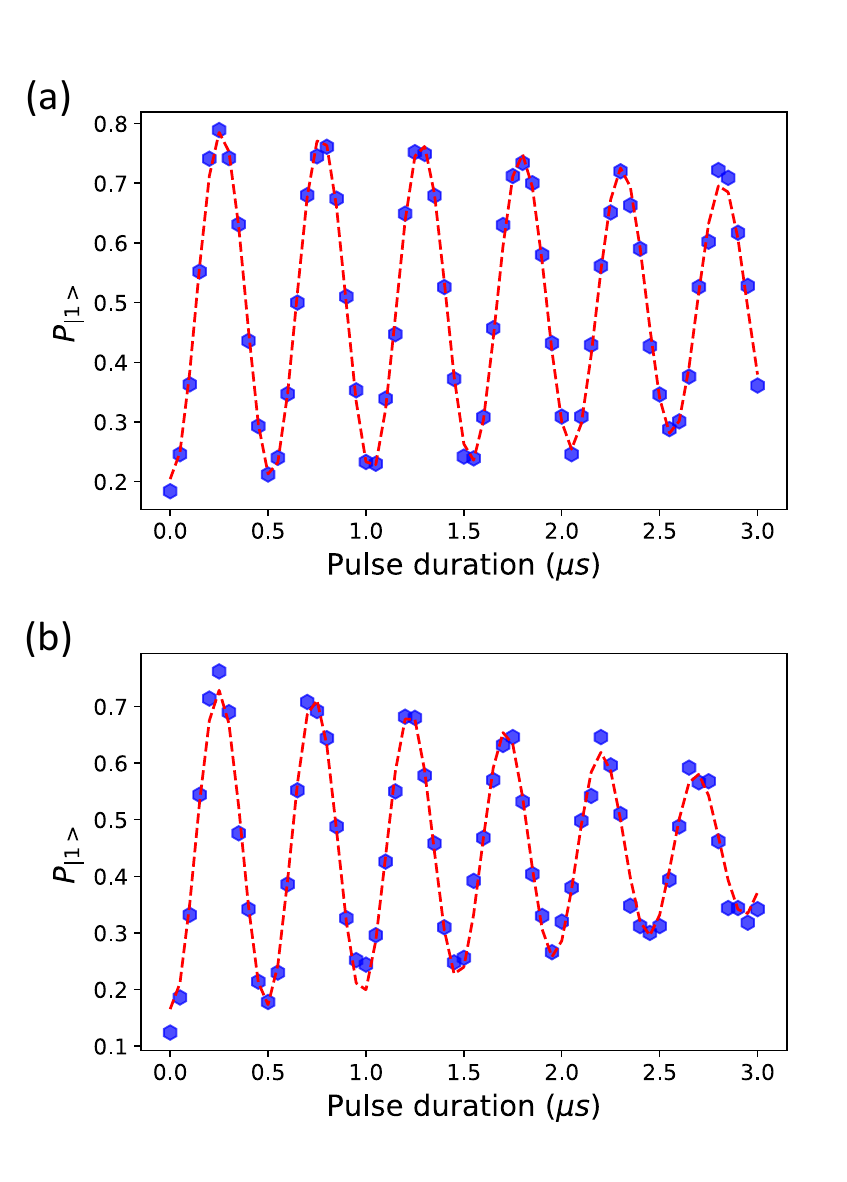}
}
\caption{
Rabi oscillations of Q2. (a) Probability that measurement of Q2 returns spin up ($\ket{1}$) as a function of the duration of the resonant microwave burst driving the qubit. (b) Analogous data for Q2 when Q1 is simultaneously being driven.
}
\label{fig:rabidecay}
\end{figure}

We here provide more information on the cross-talk effects that occur on one qubit when applying a microwave drive to the other (see also~\cite{watson2018programmable} and the supplementary information therein). First, when we perform spectroscopy on Q2 while driving Q1, we find that the frequency of Q2 shifts by of the order of 2 MHz (depending on the power applied to Q1). We compensate for this known frequency shift by shifting the drive frequency applied to Q2 when we simultaneously drive Q1. We note that a frequency shift by a known amount is not expected to contribute to decoherence. However, Fig.~\ref{fig:rabidecay} shows Rabi oscillations for both qubits in the absence and presence of an excitation to the other qubit. Clearly, when simultaneously driving Rabi oscillations on both qubits, we find a faster decay on Q2 comparing to driving Q2 by itself. The effect of simultaneous driving on Q1 is less pronounced. This is consistent with the observed effects of simultaneous driving on the measured single-qubit gate fidelities reported in the main text. The cross-talk effect on Q2 persists when the drive on Q1 is applied off-resonantly or when dot 1 is emptied. We do note that the microwave power used to drive Q1 ($\sim$20dBm) is substantially higher than that used for Q2 ($\sim$8dbm). This difference is needed to compensate for the tighter confining potential of dot 1 compared to dot 2.

\bibliography{reference}

\begin{thebibliography}{35}%
\makeatletter
\providecommand \@ifxundefined [1]{%
 \@ifx{#1\undefined}
}%
\providecommand \@ifnum [1]{%
 \ifnum #1\expandafter \@firstoftwo
 \else \expandafter \@secondoftwo
 \fi
}%
\providecommand \@ifx [1]{%
 \ifx #1\expandafter \@firstoftwo
 \else \expandafter \@secondoftwo
 \fi
}%
\providecommand \natexlab [1]{#1}%
\providecommand \enquote  [1]{``#1''}%
\providecommand \bibnamefont  [1]{#1}%
\providecommand \bibfnamefont [1]{#1}%
\providecommand \citenamefont [1]{#1}%
\providecommand \href@noop [0]{\@secondoftwo}%
\providecommand \href [0]{\begingroup \@sanitize@url \@href}%
\providecommand \@href[1]{\@@startlink{#1}\@@href}%
\providecommand \@@href[1]{\endgroup#1\@@endlink}%
\providecommand \@sanitize@url [0]{\catcode `\\12\catcode `\$12\catcode
  `\&12\catcode `\#12\catcode `\^12\catcode `\_12\catcode `\%12\relax}%
\providecommand \@@startlink[1]{}%
\providecommand \@@endlink[0]{}%
\providecommand \url  [0]{\begingroup\@sanitize@url \@url }%
\providecommand \@url [1]{\endgroup\@href {#1}{\urlprefix }}%
\providecommand \urlprefix  [0]{URL }%
\providecommand \Eprint [0]{\href }%
\providecommand \doibase [0]{http://dx.doi.org/}%
\providecommand \selectlanguage [0]{\@gobble}%
\providecommand \bibinfo  [0]{\@secondoftwo}%
\providecommand \bibfield  [0]{\@secondoftwo}%
\providecommand \translation [1]{[#1]}%
\providecommand \BibitemOpen [0]{}%
\providecommand \bibitemStop [0]{}%
\providecommand \bibitemNoStop [0]{.\EOS\space}%
\providecommand \EOS [0]{\spacefactor3000\relax}%
\providecommand \BibitemShut  [1]{\csname bibitem#1\endcsname}%
\let\auto@bib@innerbib\@empty
\bibitem [{\citenamefont {Chuang}\ and\ \citenamefont
  {Nielsen}(1997)}]{chuang1997prescription}%
  \BibitemOpen
  \bibfield  {author} {\bibinfo {author} {\bibfnamefont {Isaac~L}\ \bibnamefont
  {Chuang}}\ and\ \bibinfo {author} {\bibfnamefont {Michael~A}\ \bibnamefont
  {Nielsen}},\ }\bibfield  {title} {\enquote {\bibinfo {title} {Prescription
  for experimental determination of the dynamics of a quantum black box},}\
  }\href@noop {} {\bibfield  {journal} {\bibinfo  {journal} {Journal of Modern
  Optics}\ }\textbf {\bibinfo {volume} {44}},\ \bibinfo {pages} {2455--2467}
  (\bibinfo {year} {1997})}\BibitemShut {NoStop}%
\bibitem [{\citenamefont {O'Brien}\ \emph {et~al.}(2004)\citenamefont
  {O'Brien}, \citenamefont {Pryde}, \citenamefont {Gilchrist}, \citenamefont
  {James}, \citenamefont {Langford}, \citenamefont {Ralph},\ and\ \citenamefont
  {White}}]{o2004quantum}%
  \BibitemOpen
  \bibfield  {author} {\bibinfo {author} {\bibfnamefont {Jeremy~L}\
  \bibnamefont {O'Brien}}, \bibinfo {author} {\bibfnamefont {GJ}~\bibnamefont
  {Pryde}}, \bibinfo {author} {\bibfnamefont {Alexei}\ \bibnamefont
  {Gilchrist}}, \bibinfo {author} {\bibfnamefont {DFV}\ \bibnamefont {James}},
  \bibinfo {author} {\bibfnamefont {Nathan~K}\ \bibnamefont {Langford}},
  \bibinfo {author} {\bibfnamefont {TC}~\bibnamefont {Ralph}}, \ and\ \bibinfo
  {author} {\bibfnamefont {AG}~\bibnamefont {White}},\ }\bibfield  {title}
  {\enquote {\bibinfo {title} {Quantum process tomography of a controlled-not
  gate},}\ }\href@noop {} {\bibfield  {journal} {\bibinfo  {journal} {Physical
  review letters}\ }\textbf {\bibinfo {volume} {93}},\ \bibinfo {pages}
  {080502} (\bibinfo {year} {2004})}\BibitemShut {NoStop}%
\bibitem [{\citenamefont {Merkel}\ \emph {et~al.}(2013)\citenamefont {Merkel},
  \citenamefont {Gambetta}, \citenamefont {Smolin}, \citenamefont {Poletto},
  \citenamefont {C{\'o}rcoles}, \citenamefont {Johnson}, \citenamefont {Ryan},\
  and\ \citenamefont {Steffen}}]{merkel2013self}%
  \BibitemOpen
  \bibfield  {author} {\bibinfo {author} {\bibfnamefont {Seth~T}\ \bibnamefont
  {Merkel}}, \bibinfo {author} {\bibfnamefont {Jay~M}\ \bibnamefont
  {Gambetta}}, \bibinfo {author} {\bibfnamefont {John~A}\ \bibnamefont
  {Smolin}}, \bibinfo {author} {\bibfnamefont {Stefano}\ \bibnamefont
  {Poletto}}, \bibinfo {author} {\bibfnamefont {Antonio~D}\ \bibnamefont
  {C{\'o}rcoles}}, \bibinfo {author} {\bibfnamefont {Blake~R}\ \bibnamefont
  {Johnson}}, \bibinfo {author} {\bibfnamefont {Colm~A}\ \bibnamefont {Ryan}},
  \ and\ \bibinfo {author} {\bibfnamefont {Matthias}\ \bibnamefont {Steffen}},\
  }\bibfield  {title} {\enquote {\bibinfo {title} {Self-consistent quantum
  process tomography},}\ }\href@noop {} {\bibfield  {journal} {\bibinfo
  {journal} {Physical Review A}\ }\textbf {\bibinfo {volume} {87}},\ \bibinfo
  {pages} {062119} (\bibinfo {year} {2013})}\BibitemShut {NoStop}%
\bibitem [{\citenamefont {Emerson}\ \emph {et~al.}(2007)\citenamefont
  {Emerson}, \citenamefont {Silva}, \citenamefont {Moussa}, \citenamefont
  {Ryan}, \citenamefont {Laforest}, \citenamefont {Baugh}, \citenamefont
  {Cory},\ and\ \citenamefont {Laflamme}}]{emerson2007symmetrized}%
  \BibitemOpen
  \bibfield  {author} {\bibinfo {author} {\bibfnamefont {Joseph}\ \bibnamefont
  {Emerson}}, \bibinfo {author} {\bibfnamefont {Marcus}\ \bibnamefont {Silva}},
  \bibinfo {author} {\bibfnamefont {Osama}\ \bibnamefont {Moussa}}, \bibinfo
  {author} {\bibfnamefont {Colm}\ \bibnamefont {Ryan}}, \bibinfo {author}
  {\bibfnamefont {Martin}\ \bibnamefont {Laforest}}, \bibinfo {author}
  {\bibfnamefont {Jonathan}\ \bibnamefont {Baugh}}, \bibinfo {author}
  {\bibfnamefont {David~G}\ \bibnamefont {Cory}}, \ and\ \bibinfo {author}
  {\bibfnamefont {Raymond}\ \bibnamefont {Laflamme}},\ }\bibfield  {title}
  {\enquote {\bibinfo {title} {Symmetrized characterization of noisy quantum
  processes},}\ }\href@noop {} {\bibfield  {journal} {\bibinfo  {journal}
  {Science}\ }\textbf {\bibinfo {volume} {317}},\ \bibinfo {pages} {1893--1896}
  (\bibinfo {year} {2007})}\BibitemShut {NoStop}%
\bibitem [{\citenamefont {Knill}\ \emph {et~al.}(2008)\citenamefont {Knill},
  \citenamefont {Leibfried}, \citenamefont {Reichle}, \citenamefont {Britton},
  \citenamefont {Blakestad}, \citenamefont {Jost}, \citenamefont {Langer},
  \citenamefont {Ozeri}, \citenamefont {Seidelin},\ and\ \citenamefont
  {Wineland}}]{knill2008randomized}%
  \BibitemOpen
  \bibfield  {author} {\bibinfo {author} {\bibfnamefont {Emanuel}\ \bibnamefont
  {Knill}}, \bibinfo {author} {\bibfnamefont {D}~\bibnamefont {Leibfried}},
  \bibinfo {author} {\bibfnamefont {R}~\bibnamefont {Reichle}}, \bibinfo
  {author} {\bibfnamefont {J}~\bibnamefont {Britton}}, \bibinfo {author}
  {\bibfnamefont {RB}~\bibnamefont {Blakestad}}, \bibinfo {author}
  {\bibfnamefont {John~D}\ \bibnamefont {Jost}}, \bibinfo {author}
  {\bibfnamefont {C}~\bibnamefont {Langer}}, \bibinfo {author} {\bibfnamefont
  {R}~\bibnamefont {Ozeri}}, \bibinfo {author} {\bibfnamefont {Signe}\
  \bibnamefont {Seidelin}}, \ and\ \bibinfo {author} {\bibfnamefont {David~J}\
  \bibnamefont {Wineland}},\ }\bibfield  {title} {\enquote {\bibinfo {title}
  {Randomized benchmarking of quantum gates},}\ }\href@noop {} {\bibfield
  {journal} {\bibinfo  {journal} {Physical Review A}\ }\textbf {\bibinfo
  {volume} {77}},\ \bibinfo {pages} {012307} (\bibinfo {year}
  {2008})}\BibitemShut {NoStop}%
\bibitem [{\citenamefont {Gaebler}\ \emph {et~al.}(2012)\citenamefont
  {Gaebler}, \citenamefont {Meier}, \citenamefont {Tan}, \citenamefont
  {Bowler}, \citenamefont {Lin}, \citenamefont {Hanneke}, \citenamefont {Jost},
  \citenamefont {Home}, \citenamefont {Knill}, \citenamefont {Leibfried},\ and\
  \citenamefont {Wineland}}]{gaebler2012randomized}%
  \BibitemOpen
  \bibfield  {author} {\bibinfo {author} {\bibfnamefont {John~P}\ \bibnamefont
  {Gaebler}}, \bibinfo {author} {\bibfnamefont {Adam~M}\ \bibnamefont {Meier}},
  \bibinfo {author} {\bibfnamefont {Ting~Rei}\ \bibnamefont {Tan}}, \bibinfo
  {author} {\bibfnamefont {Ryan}\ \bibnamefont {Bowler}}, \bibinfo {author}
  {\bibfnamefont {Yiheng}\ \bibnamefont {Lin}}, \bibinfo {author}
  {\bibfnamefont {David}\ \bibnamefont {Hanneke}}, \bibinfo {author}
  {\bibfnamefont {John~D}\ \bibnamefont {Jost}}, \bibinfo {author}
  {\bibfnamefont {JP}~\bibnamefont {Home}}, \bibinfo {author} {\bibfnamefont
  {Emanuel}\ \bibnamefont {Knill}}, \bibinfo {author} {\bibfnamefont
  {Dietrich}\ \bibnamefont {Leibfried}}, \ and\ \bibinfo {author}
  {\bibfnamefont {David~J}\ \bibnamefont {Wineland}},\ }\bibfield  {title}
  {\enquote {\bibinfo {title} {Randomized benchmarking of multiqubit gates},}\
  }\href@noop {} {\bibfield  {journal} {\bibinfo  {journal} {Physical review
  letters}\ }\textbf {\bibinfo {volume} {108}},\ \bibinfo {pages} {260503}
  (\bibinfo {year} {2012})}\BibitemShut {NoStop}%
\bibitem [{\citenamefont {Chow}\ \emph {et~al.}(2009)\citenamefont {Chow},
  \citenamefont {Gambetta}, \citenamefont {Tornberg}, \citenamefont {Koch},
  \citenamefont {Bishop}, \citenamefont {Houck}, \citenamefont {Johnson},
  \citenamefont {Frunzio}, \citenamefont {Girvin},\ and\ \citenamefont
  {Schoelkopf}}]{chow2009randomized}%
  \BibitemOpen
  \bibfield  {author} {\bibinfo {author} {\bibfnamefont {JM}~\bibnamefont
  {Chow}}, \bibinfo {author} {\bibfnamefont {Jay~M}\ \bibnamefont {Gambetta}},
  \bibinfo {author} {\bibfnamefont {Lars}\ \bibnamefont {Tornberg}}, \bibinfo
  {author} {\bibfnamefont {Jens}\ \bibnamefont {Koch}}, \bibinfo {author}
  {\bibfnamefont {Lev~S}\ \bibnamefont {Bishop}}, \bibinfo {author}
  {\bibfnamefont {Andrew~A}\ \bibnamefont {Houck}}, \bibinfo {author}
  {\bibfnamefont {BR}~\bibnamefont {Johnson}}, \bibinfo {author} {\bibfnamefont
  {L}~\bibnamefont {Frunzio}}, \bibinfo {author} {\bibfnamefont {Steven~M}\
  \bibnamefont {Girvin}}, \ and\ \bibinfo {author} {\bibfnamefont {Robert~J}\
  \bibnamefont {Schoelkopf}},\ }\bibfield  {title} {\enquote {\bibinfo {title}
  {Randomized benchmarking and process tomography for gate errors in a
  solid-state qubit},}\ }\href@noop {} {\bibfield  {journal} {\bibinfo
  {journal} {Physical review letters}\ }\textbf {\bibinfo {volume} {102}},\
  \bibinfo {pages} {090502} (\bibinfo {year} {2009})}\BibitemShut {NoStop}%
\bibitem [{\citenamefont {Magesan}\ \emph
  {et~al.}(2012{\natexlab{a}})\citenamefont {Magesan}, \citenamefont
  {Gambetta}, \citenamefont {Johnson}, \citenamefont {Ryan}, \citenamefont
  {Chow}, \citenamefont {Merkel}, \citenamefont {Da~Silva}, \citenamefont
  {Keefe}, \citenamefont {Rothwell}, \citenamefont {Ohki},\ and\ \citenamefont
  {Ketchen}}]{magesan2012efficient}%
  \BibitemOpen
  \bibfield  {author} {\bibinfo {author} {\bibfnamefont {Easwar}\ \bibnamefont
  {Magesan}}, \bibinfo {author} {\bibfnamefont {Jay~M}\ \bibnamefont
  {Gambetta}}, \bibinfo {author} {\bibfnamefont {Blake~R}\ \bibnamefont
  {Johnson}}, \bibinfo {author} {\bibfnamefont {Colm~A}\ \bibnamefont {Ryan}},
  \bibinfo {author} {\bibfnamefont {Jerry~M}\ \bibnamefont {Chow}}, \bibinfo
  {author} {\bibfnamefont {Seth~T}\ \bibnamefont {Merkel}}, \bibinfo {author}
  {\bibfnamefont {Marcus~P}\ \bibnamefont {Da~Silva}}, \bibinfo {author}
  {\bibfnamefont {George~A}\ \bibnamefont {Keefe}}, \bibinfo {author}
  {\bibfnamefont {Mary~B}\ \bibnamefont {Rothwell}}, \bibinfo {author}
  {\bibfnamefont {Thomas~A}\ \bibnamefont {Ohki}}, \ and\ \bibinfo {author}
  {\bibfnamefont {Mark~B}\ \bibnamefont {Ketchen}},\ }\bibfield  {title}
  {\enquote {\bibinfo {title} {Efficient measurement of quantum gate error by
  interleaved randomized benchmarking},}\ }\href@noop {} {\bibfield  {journal}
  {\bibinfo  {journal} {Physical review letters}\ }\textbf {\bibinfo {volume}
  {109}},\ \bibinfo {pages} {080505} (\bibinfo {year}
  {2012}{\natexlab{a}})}\BibitemShut {NoStop}%
\bibitem [{\citenamefont {Gambetta}\ \emph {et~al.}(2012)\citenamefont
  {Gambetta}, \citenamefont {C{\'o}rcoles}, \citenamefont {Merkel},
  \citenamefont {Johnson}, \citenamefont {Smolin}, \citenamefont {Chow},
  \citenamefont {Ryan}, \citenamefont {Rigetti}, \citenamefont {Poletto},
  \citenamefont {Ohki}, \citenamefont {Ketchen},\ and\ \citenamefont
  {Steffen}}]{gambetta2012characterization}%
  \BibitemOpen
  \bibfield  {author} {\bibinfo {author} {\bibfnamefont {Jay~M}\ \bibnamefont
  {Gambetta}}, \bibinfo {author} {\bibfnamefont {AD}~\bibnamefont
  {C{\'o}rcoles}}, \bibinfo {author} {\bibfnamefont {Seth~T}\ \bibnamefont
  {Merkel}}, \bibinfo {author} {\bibfnamefont {Blake~R}\ \bibnamefont
  {Johnson}}, \bibinfo {author} {\bibfnamefont {John~A}\ \bibnamefont
  {Smolin}}, \bibinfo {author} {\bibfnamefont {Jerry~M}\ \bibnamefont {Chow}},
  \bibinfo {author} {\bibfnamefont {Colm~A}\ \bibnamefont {Ryan}}, \bibinfo
  {author} {\bibfnamefont {Chad}\ \bibnamefont {Rigetti}}, \bibinfo {author}
  {\bibfnamefont {S}~\bibnamefont {Poletto}}, \bibinfo {author} {\bibfnamefont
  {Thomas~A}\ \bibnamefont {Ohki}}, \bibinfo {author} {\bibfnamefont {Mark~B}\
  \bibnamefont {Ketchen}}, \ and\ \bibinfo {author} {\bibfnamefont
  {M}~\bibnamefont {Steffen}},\ }\bibfield  {title} {\enquote {\bibinfo {title}
  {Characterization of addressability by simultaneous randomized
  benchmarking},}\ }\href@noop {} {\bibfield  {journal} {\bibinfo  {journal}
  {Physical review letters}\ }\textbf {\bibinfo {volume} {109}},\ \bibinfo
  {pages} {240504} (\bibinfo {year} {2012})}\BibitemShut {NoStop}%
\bibitem [{\citenamefont {C{\'o}rcoles}\ \emph {et~al.}(2013)\citenamefont
  {C{\'o}rcoles}, \citenamefont {Gambetta}, \citenamefont {Chow}, \citenamefont
  {Smolin}, \citenamefont {Ware}, \citenamefont {Strand}, \citenamefont
  {Plourde},\ and\ \citenamefont {Steffen}}]{corcoles2013process}%
  \BibitemOpen
  \bibfield  {author} {\bibinfo {author} {\bibfnamefont {Antonio~D}\
  \bibnamefont {C{\'o}rcoles}}, \bibinfo {author} {\bibfnamefont {Jay~M}\
  \bibnamefont {Gambetta}}, \bibinfo {author} {\bibfnamefont {Jerry~M}\
  \bibnamefont {Chow}}, \bibinfo {author} {\bibfnamefont {John~A}\ \bibnamefont
  {Smolin}}, \bibinfo {author} {\bibfnamefont {Matthew}\ \bibnamefont {Ware}},
  \bibinfo {author} {\bibfnamefont {Joel}\ \bibnamefont {Strand}}, \bibinfo
  {author} {\bibfnamefont {Britton~LT}\ \bibnamefont {Plourde}}, \ and\
  \bibinfo {author} {\bibfnamefont {Matthias}\ \bibnamefont {Steffen}},\
  }\bibfield  {title} {\enquote {\bibinfo {title} {Process verification of
  two-qubit quantum gates by randomized benchmarking},}\ }\href@noop {}
  {\bibfield  {journal} {\bibinfo  {journal} {Physical Review A}\ }\textbf
  {\bibinfo {volume} {87}},\ \bibinfo {pages} {030301} (\bibinfo {year}
  {2013})}\BibitemShut {NoStop}%
\bibitem [{\citenamefont {Dugas}\ \emph {et~al.}(2016)\citenamefont {Dugas},
  \citenamefont {Wallman},\ and\ \citenamefont
  {Emerson}}]{dugas2016efficiently}%
  \BibitemOpen
  \bibfield  {author} {\bibinfo {author} {\bibfnamefont {Arnaud~C}\
  \bibnamefont {Dugas}}, \bibinfo {author} {\bibfnamefont {Joel~J}\
  \bibnamefont {Wallman}}, \ and\ \bibinfo {author} {\bibfnamefont {Joseph}\
  \bibnamefont {Emerson}},\ }\bibfield  {title} {\enquote {\bibinfo {title}
  {Efficiently characterizing the total error in quantum circuits},}\
  }\href@noop {} {\bibfield  {journal} {\bibinfo  {journal} {arXiv preprint
  arXiv:1610.05296}\ } (\bibinfo {year} {2016})}\BibitemShut {NoStop}%
\bibitem [{\citenamefont {Chen}\ \emph {et~al.}(2014)\citenamefont {Chen},
  \citenamefont {Neill}, \citenamefont {Roushan}, \citenamefont {Leung},
  \citenamefont {Fang}, \citenamefont {Barends}, \citenamefont {Kelly},
  \citenamefont {Campbell}, \citenamefont {Chen}, \citenamefont {Chiaro},
  \citenamefont {Dunsworth}, \citenamefont {Jeffrey}, \citenamefont {Megrant},
  \citenamefont {Mutus}, \citenamefont {O’Malley}, \citenamefont {Quintana},
  \citenamefont {Sank}, \citenamefont {Vainsencher}, \citenamefont {Wenner},
  \citenamefont {White}, \citenamefont {Geller}, \citenamefont {Cleland},\ and\
  \citenamefont {Martinis}}]{chen2014qubit}%
  \BibitemOpen
  \bibfield  {author} {\bibinfo {author} {\bibfnamefont {Yu}~\bibnamefont
  {Chen}}, \bibinfo {author} {\bibfnamefont {C}~\bibnamefont {Neill}}, \bibinfo
  {author} {\bibfnamefont {P}~\bibnamefont {Roushan}}, \bibinfo {author}
  {\bibfnamefont {N}~\bibnamefont {Leung}}, \bibinfo {author} {\bibfnamefont
  {M}~\bibnamefont {Fang}}, \bibinfo {author} {\bibfnamefont {R}~\bibnamefont
  {Barends}}, \bibinfo {author} {\bibfnamefont {J}~\bibnamefont {Kelly}},
  \bibinfo {author} {\bibfnamefont {B}~\bibnamefont {Campbell}}, \bibinfo
  {author} {\bibfnamefont {Z}~\bibnamefont {Chen}}, \bibinfo {author}
  {\bibfnamefont {B}~\bibnamefont {Chiaro}}, \bibinfo {author} {\bibfnamefont
  {A}~\bibnamefont {Dunsworth}}, \bibinfo {author} {\bibfnamefont
  {E}~\bibnamefont {Jeffrey}}, \bibinfo {author} {\bibfnamefont
  {A}~\bibnamefont {Megrant}}, \bibinfo {author} {\bibfnamefont
  {JY}~\bibnamefont {Mutus}}, \bibinfo {author} {\bibfnamefont
  {JJ}~\bibnamefont {O’Malley}}, \bibinfo {author} {\bibfnamefont
  {CM}~\bibnamefont {Quintana}}, \bibinfo {author} {\bibfnamefont
  {D}~\bibnamefont {Sank}}, \bibinfo {author} {\bibfnamefont {A}~\bibnamefont
  {Vainsencher}}, \bibinfo {author} {\bibfnamefont {J}~\bibnamefont {Wenner}},
  \bibinfo {author} {\bibfnamefont {TC}~\bibnamefont {White}}, \bibinfo
  {author} {\bibfnamefont {Michael~R}\ \bibnamefont {Geller}}, \bibinfo
  {author} {\bibfnamefont {AN}~\bibnamefont {Cleland}}, \ and\ \bibinfo
  {author} {\bibfnamefont {John~M}\ \bibnamefont {Martinis}},\ }\bibfield
  {title} {\enquote {\bibinfo {title} {Qubit architecture with high coherence
  and fast tunable coupling},}\ }\href@noop {} {\bibfield  {journal} {\bibinfo
  {journal} {Physical review letters}\ }\textbf {\bibinfo {volume} {113}},\
  \bibinfo {pages} {220502} (\bibinfo {year} {2014})}\BibitemShut {NoStop}%
\bibitem [{\citenamefont {Casparis}\ \emph {et~al.}(2016)\citenamefont
  {Casparis}, \citenamefont {Larsen}, \citenamefont {Olsen}, \citenamefont
  {Kuemmeth}, \citenamefont {Krogstrup}, \citenamefont {Nyg{\aa}rd},
  \citenamefont {Petersson},\ and\ \citenamefont
  {Marcus}}]{casparis2016gatemon}%
  \BibitemOpen
  \bibfield  {author} {\bibinfo {author} {\bibfnamefont {Lucas}\ \bibnamefont
  {Casparis}}, \bibinfo {author} {\bibfnamefont {TW}~\bibnamefont {Larsen}},
  \bibinfo {author} {\bibfnamefont {MS}~\bibnamefont {Olsen}}, \bibinfo
  {author} {\bibfnamefont {F}~\bibnamefont {Kuemmeth}}, \bibinfo {author}
  {\bibfnamefont {P}~\bibnamefont {Krogstrup}}, \bibinfo {author}
  {\bibfnamefont {Jesper}\ \bibnamefont {Nyg{\aa}rd}}, \bibinfo {author}
  {\bibfnamefont {KD}~\bibnamefont {Petersson}}, \ and\ \bibinfo {author}
  {\bibfnamefont {CM}~\bibnamefont {Marcus}},\ }\bibfield  {title} {\enquote
  {\bibinfo {title} {Gatemon benchmarking and two-qubit operations},}\
  }\href@noop {} {\bibfield  {journal} {\bibinfo  {journal} {Physical review
  letters}\ }\textbf {\bibinfo {volume} {116}},\ \bibinfo {pages} {150505}
  (\bibinfo {year} {2016})}\BibitemShut {NoStop}%
\bibitem [{\citenamefont {Shulman}\ \emph {et~al.}(2012)\citenamefont
  {Shulman}, \citenamefont {Dial}, \citenamefont {Harvey}, \citenamefont
  {Bluhm}, \citenamefont {Umansky},\ and\ \citenamefont
  {Yacoby}}]{shulman2012demonstration}%
  \BibitemOpen
  \bibfield  {author} {\bibinfo {author} {\bibfnamefont {Michael~Dean}\
  \bibnamefont {Shulman}}, \bibinfo {author} {\bibfnamefont {Oliver~E}\
  \bibnamefont {Dial}}, \bibinfo {author} {\bibfnamefont {Shannon~Pasca}\
  \bibnamefont {Harvey}}, \bibinfo {author} {\bibfnamefont {Hendrik}\
  \bibnamefont {Bluhm}}, \bibinfo {author} {\bibfnamefont {Vladimir}\
  \bibnamefont {Umansky}}, \ and\ \bibinfo {author} {\bibfnamefont {Amir}\
  \bibnamefont {Yacoby}},\ }\bibfield  {title} {\enquote {\bibinfo {title}
  {Demonstration of entanglement of electrostatically coupled singlet-triplet
  qubits},}\ }\href@noop {} {\bibfield  {journal} {\bibinfo  {journal}
  {Science}\ }\textbf {\bibinfo {volume} {336}},\ \bibinfo {pages} {202--205}
  (\bibinfo {year} {2012})}\BibitemShut {NoStop}%
\bibitem [{\citenamefont {Watson}\ \emph {et~al.}(2018)\citenamefont {Watson},
  \citenamefont {Philips}, \citenamefont {Kawakami}, \citenamefont {Ward},
  \citenamefont {Scarlino}, \citenamefont {Veldhorst}, \citenamefont {Savage},
  \citenamefont {Lagally}, \citenamefont {Friesen}, \citenamefont
  {Coppersmith}, \citenamefont {Eriksson},\ and\ \citenamefont
  {Vandersypen}}]{watson2018programmable}%
  \BibitemOpen
  \bibfield  {author} {\bibinfo {author} {\bibfnamefont {TF}~\bibnamefont
  {Watson}}, \bibinfo {author} {\bibfnamefont {SGJ}\ \bibnamefont {Philips}},
  \bibinfo {author} {\bibfnamefont {Erika}\ \bibnamefont {Kawakami}}, \bibinfo
  {author} {\bibfnamefont {DR}~\bibnamefont {Ward}}, \bibinfo {author}
  {\bibfnamefont {Pasquale}\ \bibnamefont {Scarlino}}, \bibinfo {author}
  {\bibfnamefont {Menno}\ \bibnamefont {Veldhorst}}, \bibinfo {author}
  {\bibfnamefont {DE}~\bibnamefont {Savage}}, \bibinfo {author} {\bibfnamefont
  {MG}~\bibnamefont {Lagally}}, \bibinfo {author} {\bibfnamefont {Mark}\
  \bibnamefont {Friesen}}, \bibinfo {author} {\bibfnamefont {SN}~\bibnamefont
  {Coppersmith}}, \bibinfo {author} {\bibfnamefont {MA}~\bibnamefont
  {Eriksson}}, \ and\ \bibinfo {author} {\bibfnamefont {LMK}\ \bibnamefont
  {Vandersypen}},\ }\bibfield  {title} {\enquote {\bibinfo {title} {A
  programmable two-qubit quantum processor in silicon},}\ }\href@noop {}
  {\bibfield  {journal} {\bibinfo  {journal} {Nature}\ } (\bibinfo {year}
  {2018})}\BibitemShut {NoStop}%
\bibitem [{\citenamefont {Zajac}\ \emph {et~al.}(2018)\citenamefont {Zajac},
  \citenamefont {Sigillito}, \citenamefont {Russ}, \citenamefont {Borjans},
  \citenamefont {Taylor}, \citenamefont {Burkard},\ and\ \citenamefont
  {Petta}}]{zajac2018resonantly}%
  \BibitemOpen
  \bibfield  {author} {\bibinfo {author} {\bibfnamefont {David~M}\ \bibnamefont
  {Zajac}}, \bibinfo {author} {\bibfnamefont {Anthony~J}\ \bibnamefont
  {Sigillito}}, \bibinfo {author} {\bibfnamefont {Maximilian}\ \bibnamefont
  {Russ}}, \bibinfo {author} {\bibfnamefont {Felix}\ \bibnamefont {Borjans}},
  \bibinfo {author} {\bibfnamefont {Jacob~M}\ \bibnamefont {Taylor}}, \bibinfo
  {author} {\bibfnamefont {Guido}\ \bibnamefont {Burkard}}, \ and\ \bibinfo
  {author} {\bibfnamefont {Jason~R}\ \bibnamefont {Petta}},\ }\bibfield
  {title} {\enquote {\bibinfo {title} {Resonantly driven cnot gate for electron
  spins},}\ }\href@noop {} {\bibfield  {journal} {\bibinfo  {journal}
  {Science}\ }\textbf {\bibinfo {volume} {359}},\ \bibinfo {pages} {439--442}
  (\bibinfo {year} {2018})}\BibitemShut {NoStop}%
\bibitem [{\citenamefont {Huang}\ \emph {et~al.}(2018)\citenamefont {Huang},
  \citenamefont {Yang}, \citenamefont {Chan}, \citenamefont {Tanttu},
  \citenamefont {Hensen}, \citenamefont {Leon}, \citenamefont {Fogarty},
  \citenamefont {Hwang}, \citenamefont {Hudson}, \citenamefont {Itoh},
  \citenamefont {Morello}, \citenamefont {Laucht},\ and\ \citenamefont
  {Dzurak}}]{huang2018fidelity}%
  \BibitemOpen
  \bibfield  {author} {\bibinfo {author} {\bibfnamefont {W}~\bibnamefont
  {Huang}}, \bibinfo {author} {\bibfnamefont {CH}~\bibnamefont {Yang}},
  \bibinfo {author} {\bibfnamefont {KW}~\bibnamefont {Chan}}, \bibinfo {author}
  {\bibfnamefont {T}~\bibnamefont {Tanttu}}, \bibinfo {author} {\bibfnamefont
  {B}~\bibnamefont {Hensen}}, \bibinfo {author} {\bibfnamefont {RCC}\
  \bibnamefont {Leon}}, \bibinfo {author} {\bibfnamefont {MA}~\bibnamefont
  {Fogarty}}, \bibinfo {author} {\bibfnamefont {JCC}\ \bibnamefont {Hwang}},
  \bibinfo {author} {\bibfnamefont {FE}~\bibnamefont {Hudson}}, \bibinfo
  {author} {\bibfnamefont {KM}~\bibnamefont {Itoh}}, \bibinfo {author}
  {\bibfnamefont {A}~\bibnamefont {Morello}}, \bibinfo {author} {\bibfnamefont
  {A}~\bibnamefont {Laucht}}, \ and\ \bibinfo {author} {\bibfnamefont
  {AS}~\bibnamefont {Dzurak}},\ }\bibfield  {title} {\enquote {\bibinfo {title}
  {Fidelity benchmarks for two-qubit gates in silicon},}\ }\href@noop {}
  {\bibfield  {journal} {\bibinfo  {journal} {arXiv preprint arXiv:1805.05027}\
  } (\bibinfo {year} {2018})}\BibitemShut {NoStop}%
\bibitem [{\citenamefont {Helsen}\ \emph {et~al.}(2018)\citenamefont {Helsen},
  \citenamefont {Xue}, \citenamefont {Vandersypen},\ and\ \citenamefont
  {Wehner}}]{helsen2018new}%
  \BibitemOpen
  \bibfield  {author} {\bibinfo {author} {\bibfnamefont {Jonas}\ \bibnamefont
  {Helsen}}, \bibinfo {author} {\bibfnamefont {Xiao}\ \bibnamefont {Xue}},
  \bibinfo {author} {\bibfnamefont {Lieven~MK}\ \bibnamefont {Vandersypen}}, \
  and\ \bibinfo {author} {\bibfnamefont {Stephanie}\ \bibnamefont {Wehner}},\
  }\bibfield  {title} {\enquote {\bibinfo {title} {A new class of efficient
  randomized benchmarking protocols},}\ }\href@noop {} {\bibfield  {journal}
  {\bibinfo  {journal} {arXiv preprint arXiv:1806.02048}\ } (\bibinfo {year}
  {2018})}\BibitemShut {NoStop}%
\bibitem [{\citenamefont {Zwanenburg}\ \emph {et~al.}(2013)\citenamefont
  {Zwanenburg}, \citenamefont {Dzurak}, \citenamefont {Morello}, \citenamefont
  {Simmons}, \citenamefont {Hollenberg}, \citenamefont {Klimeck}, \citenamefont
  {Rogge}, \citenamefont {Coppersmith},\ and\ \citenamefont
  {Eriksson}}]{zwanenburg2013silicon}%
  \BibitemOpen
  \bibfield  {author} {\bibinfo {author} {\bibfnamefont {Floris~A}\
  \bibnamefont {Zwanenburg}}, \bibinfo {author} {\bibfnamefont {Andrew~S}\
  \bibnamefont {Dzurak}}, \bibinfo {author} {\bibfnamefont {Andrea}\
  \bibnamefont {Morello}}, \bibinfo {author} {\bibfnamefont {Michelle~Y}\
  \bibnamefont {Simmons}}, \bibinfo {author} {\bibfnamefont {Lloyd~CL}\
  \bibnamefont {Hollenberg}}, \bibinfo {author} {\bibfnamefont {Gerhard}\
  \bibnamefont {Klimeck}}, \bibinfo {author} {\bibfnamefont {Sven}\
  \bibnamefont {Rogge}}, \bibinfo {author} {\bibfnamefont {Susan~N}\
  \bibnamefont {Coppersmith}}, \ and\ \bibinfo {author} {\bibfnamefont
  {Mark~A}\ \bibnamefont {Eriksson}},\ }\bibfield  {title} {\enquote {\bibinfo
  {title} {Silicon quantum electronics},}\ }\href@noop {} {\bibfield  {journal}
  {\bibinfo  {journal} {Reviews of modern physics}\ }\textbf {\bibinfo {volume}
  {85}},\ \bibinfo {pages} {961} (\bibinfo {year} {2013})}\BibitemShut
  {NoStop}%
\bibitem [{\citenamefont {Vandersypen}\ \emph {et~al.}(2017)\citenamefont
  {Vandersypen}, \citenamefont {Bluhm}, \citenamefont {Clarke}, \citenamefont
  {Dzurak}, \citenamefont {Ishihara}, \citenamefont {Morello}, \citenamefont
  {Reilly}, \citenamefont {Schreiber},\ and\ \citenamefont
  {Veldhorst}}]{vandersypen2017interfacing}%
  \BibitemOpen
  \bibfield  {author} {\bibinfo {author} {\bibfnamefont {LMK}\ \bibnamefont
  {Vandersypen}}, \bibinfo {author} {\bibfnamefont {H}~\bibnamefont {Bluhm}},
  \bibinfo {author} {\bibfnamefont {JS}~\bibnamefont {Clarke}}, \bibinfo
  {author} {\bibfnamefont {AS}~\bibnamefont {Dzurak}}, \bibinfo {author}
  {\bibfnamefont {R}~\bibnamefont {Ishihara}}, \bibinfo {author} {\bibfnamefont
  {A}~\bibnamefont {Morello}}, \bibinfo {author} {\bibfnamefont
  {DJ}~\bibnamefont {Reilly}}, \bibinfo {author} {\bibfnamefont
  {LR}~\bibnamefont {Schreiber}}, \ and\ \bibinfo {author} {\bibfnamefont
  {M}~\bibnamefont {Veldhorst}},\ }\bibfield  {title} {\enquote {\bibinfo
  {title} {Interfacing spin qubits in quantum dots and donors—hot, dense, and
  coherent},}\ }\href@noop {} {\bibfield  {journal} {\bibinfo  {journal} {npj
  Quantum Information}\ }\textbf {\bibinfo {volume} {3}},\ \bibinfo {pages}
  {34} (\bibinfo {year} {2017})}\BibitemShut {NoStop}%
\bibitem [{\citenamefont {Pioro-Ladriere}\ \emph {et~al.}(2008)\citenamefont
  {Pioro-Ladriere}, \citenamefont {Obata}, \citenamefont {Tokura},
  \citenamefont {Shin}, \citenamefont {Kubo}, \citenamefont {Yoshida},
  \citenamefont {Taniyama},\ and\ \citenamefont
  {Tarucha}}]{pioro2008electrically}%
  \BibitemOpen
  \bibfield  {author} {\bibinfo {author} {\bibfnamefont {M}~\bibnamefont
  {Pioro-Ladriere}}, \bibinfo {author} {\bibfnamefont {T}~\bibnamefont
  {Obata}}, \bibinfo {author} {\bibfnamefont {Y}~\bibnamefont {Tokura}},
  \bibinfo {author} {\bibfnamefont {Y-S}\ \bibnamefont {Shin}}, \bibinfo
  {author} {\bibfnamefont {T}~\bibnamefont {Kubo}}, \bibinfo {author}
  {\bibfnamefont {K}~\bibnamefont {Yoshida}}, \bibinfo {author} {\bibfnamefont
  {T}~\bibnamefont {Taniyama}}, \ and\ \bibinfo {author} {\bibfnamefont
  {S}~\bibnamefont {Tarucha}},\ }\bibfield  {title} {\enquote {\bibinfo {title}
  {Electrically driven single-electron spin resonance in a slanting zeeman
  field},}\ }\href@noop {} {\bibfield  {journal} {\bibinfo  {journal} {Nature
  Physics}\ }\textbf {\bibinfo {volume} {4}},\ \bibinfo {pages} {776} (\bibinfo
  {year} {2008})}\BibitemShut {NoStop}%
\bibitem [{\citenamefont {Vandersypen}\ and\ \citenamefont
  {Chuang}(2005)}]{vandersypen2005nmr}%
  \BibitemOpen
  \bibfield  {author} {\bibinfo {author} {\bibfnamefont {Lieven~MK}\
  \bibnamefont {Vandersypen}}\ and\ \bibinfo {author} {\bibfnamefont {Isaac~L}\
  \bibnamefont {Chuang}},\ }\bibfield  {title} {\enquote {\bibinfo {title} {Nmr
  techniques for quantum control and computation},}\ }\href@noop {} {\bibfield
  {journal} {\bibinfo  {journal} {Reviews of modern physics}\ }\textbf
  {\bibinfo {volume} {76}},\ \bibinfo {pages} {1037} (\bibinfo {year}
  {2005})}\BibitemShut {NoStop}%
\bibitem [{\citenamefont {Meunier}\ \emph {et~al.}(2011)\citenamefont
  {Meunier}, \citenamefont {Calado},\ and\ \citenamefont
  {Vandersypen}}]{meunier2011efficient}%
  \BibitemOpen
  \bibfield  {author} {\bibinfo {author} {\bibfnamefont {Tristan}\ \bibnamefont
  {Meunier}}, \bibinfo {author} {\bibfnamefont {VE}~\bibnamefont {Calado}}, \
  and\ \bibinfo {author} {\bibfnamefont {LMK}\ \bibnamefont {Vandersypen}},\
  }\bibfield  {title} {\enquote {\bibinfo {title} {Efficient controlled-phase
  gate for single-spin qubits in quantum dots},}\ }\href@noop {} {\bibfield
  {journal} {\bibinfo  {journal} {Physical Review B}\ }\textbf {\bibinfo
  {volume} {83}},\ \bibinfo {pages} {121403} (\bibinfo {year}
  {2011})}\BibitemShut {NoStop}%
\bibitem [{\citenamefont {Veldhorst}\ \emph {et~al.}(2015)\citenamefont
  {Veldhorst}, \citenamefont {Yang}, \citenamefont {Hwang}, \citenamefont
  {Huang}, \citenamefont {Dehollain}, \citenamefont {Muhonen}, \citenamefont
  {Simmons}, \citenamefont {Laucht}, \citenamefont {Hudson}, \citenamefont
  {Itoh}, \citenamefont {Morello},\ and\ \citenamefont
  {Dzurak}}]{veldhorst2015two}%
  \BibitemOpen
  \bibfield  {author} {\bibinfo {author} {\bibfnamefont {Menno}\ \bibnamefont
  {Veldhorst}}, \bibinfo {author} {\bibfnamefont {CH}~\bibnamefont {Yang}},
  \bibinfo {author} {\bibfnamefont {JCC}\ \bibnamefont {Hwang}}, \bibinfo
  {author} {\bibfnamefont {W}~\bibnamefont {Huang}}, \bibinfo {author}
  {\bibfnamefont {JP}~\bibnamefont {Dehollain}}, \bibinfo {author}
  {\bibfnamefont {JT}~\bibnamefont {Muhonen}}, \bibinfo {author} {\bibfnamefont
  {S}~\bibnamefont {Simmons}}, \bibinfo {author} {\bibfnamefont
  {A}~\bibnamefont {Laucht}}, \bibinfo {author} {\bibfnamefont
  {FE}~\bibnamefont {Hudson}}, \bibinfo {author} {\bibfnamefont {Kohei~M}\
  \bibnamefont {Itoh}}, \bibinfo {author} {\bibfnamefont {A}~\bibnamefont
  {Morello}}, \ and\ \bibinfo {author} {\bibfnamefont {AS}~\bibnamefont
  {Dzurak}},\ }\bibfield  {title} {\enquote {\bibinfo {title} {A two-qubit
  logic gate in silicon},}\ }\href@noop {} {\bibfield  {journal} {\bibinfo
  {journal} {Nature}\ }\textbf {\bibinfo {volume} {526}},\ \bibinfo {pages}
  {410} (\bibinfo {year} {2015})}\BibitemShut {NoStop}%
\bibitem [{\citenamefont {Elzerman}\ \emph {et~al.}(2004)\citenamefont
  {Elzerman}, \citenamefont {Hanson}, \citenamefont {Van~Beveren},
  \citenamefont {Witkamp}, \citenamefont {Vandersypen},\ and\ \citenamefont
  {Kouwenhoven}}]{elzerman2004single}%
  \BibitemOpen
  \bibfield  {author} {\bibinfo {author} {\bibfnamefont {JM}~\bibnamefont
  {Elzerman}}, \bibinfo {author} {\bibfnamefont {R}~\bibnamefont {Hanson}},
  \bibinfo {author} {\bibfnamefont {LH~Willems}\ \bibnamefont {Van~Beveren}},
  \bibinfo {author} {\bibfnamefont {B}~\bibnamefont {Witkamp}}, \bibinfo
  {author} {\bibfnamefont {LMK}\ \bibnamefont {Vandersypen}}, \ and\ \bibinfo
  {author} {\bibfnamefont {Leo~P}\ \bibnamefont {Kouwenhoven}},\ }\bibfield
  {title} {\enquote {\bibinfo {title} {Single-shot read-out of an individual
  electron spin in a quantum dot},}\ }\href@noop {} {\bibfield  {journal}
  {\bibinfo  {journal} {Nature}\ }\textbf {\bibinfo {volume} {430}},\ \bibinfo
  {pages} {431} (\bibinfo {year} {2004})}\BibitemShut {NoStop}%
\bibitem [{\citenamefont {Srinivasa}\ \emph {et~al.}(2013)\citenamefont
  {Srinivasa}, \citenamefont {Nowack}, \citenamefont {Shafiei}, \citenamefont
  {Vandersypen},\ and\ \citenamefont {Taylor}}]{srinivasa2013simultaneous}%
  \BibitemOpen
  \bibfield  {author} {\bibinfo {author} {\bibfnamefont {Vanita}\ \bibnamefont
  {Srinivasa}}, \bibinfo {author} {\bibfnamefont {Katja~C}\ \bibnamefont
  {Nowack}}, \bibinfo {author} {\bibfnamefont {Mohammad}\ \bibnamefont
  {Shafiei}}, \bibinfo {author} {\bibfnamefont {LMK}\ \bibnamefont
  {Vandersypen}}, \ and\ \bibinfo {author} {\bibfnamefont {Jacob~M}\
  \bibnamefont {Taylor}},\ }\bibfield  {title} {\enquote {\bibinfo {title}
  {Simultaneous spin-charge relaxation in double quantum dots},}\ }\href@noop
  {} {\bibfield  {journal} {\bibinfo  {journal} {Physical review letters}\
  }\textbf {\bibinfo {volume} {110}},\ \bibinfo {pages} {196803} (\bibinfo
  {year} {2013})}\BibitemShut {NoStop}%
\bibitem [{\citenamefont {Magesan}\ \emph {et~al.}(2011)\citenamefont
  {Magesan}, \citenamefont {Gambetta},\ and\ \citenamefont
  {Emerson}}]{magesan2011scalable}%
  \BibitemOpen
  \bibfield  {author} {\bibinfo {author} {\bibfnamefont {Easwar}\ \bibnamefont
  {Magesan}}, \bibinfo {author} {\bibfnamefont {Jay~M}\ \bibnamefont
  {Gambetta}}, \ and\ \bibinfo {author} {\bibfnamefont {Joseph}\ \bibnamefont
  {Emerson}},\ }\bibfield  {title} {\enquote {\bibinfo {title} {Scalable and
  robust randomized benchmarking of quantum processes},}\ }\href@noop {}
  {\bibfield  {journal} {\bibinfo  {journal} {Physical review letters}\
  }\textbf {\bibinfo {volume} {106}},\ \bibinfo {pages} {180504} (\bibinfo
  {year} {2011})}\BibitemShut {NoStop}%
\bibitem [{\citenamefont {Magesan}\ \emph
  {et~al.}(2012{\natexlab{b}})\citenamefont {Magesan}, \citenamefont
  {Gambetta},\ and\ \citenamefont {Emerson}}]{magesan2012characterizing}%
  \BibitemOpen
  \bibfield  {author} {\bibinfo {author} {\bibfnamefont {Easwar}\ \bibnamefont
  {Magesan}}, \bibinfo {author} {\bibfnamefont {Jay~M}\ \bibnamefont
  {Gambetta}}, \ and\ \bibinfo {author} {\bibfnamefont {Joseph}\ \bibnamefont
  {Emerson}},\ }\bibfield  {title} {\enquote {\bibinfo {title} {Characterizing
  quantum gates via randomized benchmarking},}\ }\href@noop {} {\bibfield
  {journal} {\bibinfo  {journal} {Physical Review A}\ }\textbf {\bibinfo
  {volume} {85}},\ \bibinfo {pages} {042311} (\bibinfo {year}
  {2012}{\natexlab{b}})}\BibitemShut {NoStop}%
\bibitem [{\citenamefont {Wallman}(2018)}]{wallman2018randomized}%
  \BibitemOpen
  \bibfield  {author} {\bibinfo {author} {\bibfnamefont {Joel~J}\ \bibnamefont
  {Wallman}},\ }\bibfield  {title} {\enquote {\bibinfo {title} {Randomized
  benchmarking with gate-dependent noise},}\ }\href@noop {} {\bibfield
  {journal} {\bibinfo  {journal} {Quantum}\ }\textbf {\bibinfo {volume} {2}},\
  \bibinfo {pages} {47} (\bibinfo {year} {2018})}\BibitemShut {NoStop}%
\bibitem [{\citenamefont {Carignan-Dugas}\ \emph {et~al.}(2018)\citenamefont
  {Carignan-Dugas}, \citenamefont {Boone}, \citenamefont {Wallman},\ and\
  \citenamefont {Emerson}}]{carignan2018randomized}%
  \BibitemOpen
  \bibfield  {author} {\bibinfo {author} {\bibfnamefont {Arnaud}\ \bibnamefont
  {Carignan-Dugas}}, \bibinfo {author} {\bibfnamefont {Kristine}\ \bibnamefont
  {Boone}}, \bibinfo {author} {\bibfnamefont {Joel~J}\ \bibnamefont {Wallman}},
  \ and\ \bibinfo {author} {\bibfnamefont {Joseph}\ \bibnamefont {Emerson}},\
  }\bibfield  {title} {\enquote {\bibinfo {title} {From randomized benchmarking
  experiments to gate-set circuit fidelity: how to interpret randomized
  benchmarking decay parameters},}\ }\href@noop {} {\bibfield  {journal}
  {\bibinfo  {journal} {New Journal of Physics}\ }\textbf {\bibinfo {volume}
  {20}},\ \bibinfo {pages} {092001} (\bibinfo {year} {2018})}\BibitemShut
  {NoStop}%
\bibitem [{\citenamefont {Proctor}\ \emph {et~al.}(2017)\citenamefont
  {Proctor}, \citenamefont {Rudinger}, \citenamefont {Young}, \citenamefont
  {Sarovar},\ and\ \citenamefont {Blume-Kohout}}]{proctor2017randomized}%
  \BibitemOpen
  \bibfield  {author} {\bibinfo {author} {\bibfnamefont {Timothy}\ \bibnamefont
  {Proctor}}, \bibinfo {author} {\bibfnamefont {Kenneth}\ \bibnamefont
  {Rudinger}}, \bibinfo {author} {\bibfnamefont {Kevin}\ \bibnamefont {Young}},
  \bibinfo {author} {\bibfnamefont {Mohan}\ \bibnamefont {Sarovar}}, \ and\
  \bibinfo {author} {\bibfnamefont {Robin}\ \bibnamefont {Blume-Kohout}},\
  }\bibfield  {title} {\enquote {\bibinfo {title} {What randomized benchmarking
  actually measures},}\ }\href@noop {} {\bibfield  {journal} {\bibinfo
  {journal} {Physical review letters}\ }\textbf {\bibinfo {volume} {119}},\
  \bibinfo {pages} {130502} (\bibinfo {year} {2017})}\BibitemShut {NoStop}%
\bibitem [{\citenamefont {Martins}\ \emph {et~al.}(2016)\citenamefont
  {Martins}, \citenamefont {Malinowski}, \citenamefont {Nissen}, \citenamefont
  {Barnes}, \citenamefont {Fallahi}, \citenamefont {Gardner}, \citenamefont
  {Manfra}, \citenamefont {Marcus},\ and\ \citenamefont
  {Kuemmeth}}]{martins2016noise}%
  \BibitemOpen
  \bibfield  {author} {\bibinfo {author} {\bibfnamefont {Frederico}\
  \bibnamefont {Martins}}, \bibinfo {author} {\bibfnamefont {Filip~K}\
  \bibnamefont {Malinowski}}, \bibinfo {author} {\bibfnamefont {Peter~D}\
  \bibnamefont {Nissen}}, \bibinfo {author} {\bibfnamefont {Edwin}\
  \bibnamefont {Barnes}}, \bibinfo {author} {\bibfnamefont {Saeed}\
  \bibnamefont {Fallahi}}, \bibinfo {author} {\bibfnamefont {Geoffrey~C}\
  \bibnamefont {Gardner}}, \bibinfo {author} {\bibfnamefont {Michael~J}\
  \bibnamefont {Manfra}}, \bibinfo {author} {\bibfnamefont {Charles~M}\
  \bibnamefont {Marcus}}, \ and\ \bibinfo {author} {\bibfnamefont {Ferdinand}\
  \bibnamefont {Kuemmeth}},\ }\bibfield  {title} {\enquote {\bibinfo {title}
  {Noise suppression using symmetric exchange gates in spin qubits},}\
  }\href@noop {} {\bibfield  {journal} {\bibinfo  {journal} {Physical review
  letters}\ }\textbf {\bibinfo {volume} {116}},\ \bibinfo {pages} {116801}
  (\bibinfo {year} {2016})}\BibitemShut {NoStop}%
\bibitem [{\citenamefont {Reed}\ \emph {et~al.}(2016)\citenamefont {Reed},
  \citenamefont {Maune}, \citenamefont {Andrews}, \citenamefont {Borselli},
  \citenamefont {Eng}, \citenamefont {Jura}, \citenamefont {Kiselev},
  \citenamefont {Ladd}, \citenamefont {Merkel}, \citenamefont {Milosavljevic},
  \citenamefont {Pritchett}, \citenamefont {Rakher}, \citenamefont {Ross},
  \citenamefont {Schmitz}, \citenamefont {Smith}, \citenamefont {Wright},
  \citenamefont {Gyure},\ and\ \citenamefont {Hunter}}]{reed2016reduced}%
  \BibitemOpen
  \bibfield  {author} {\bibinfo {author} {\bibfnamefont {MD}~\bibnamefont
  {Reed}}, \bibinfo {author} {\bibfnamefont {BM}~\bibnamefont {Maune}},
  \bibinfo {author} {\bibfnamefont {RW}~\bibnamefont {Andrews}}, \bibinfo
  {author} {\bibfnamefont {MG}~\bibnamefont {Borselli}}, \bibinfo {author}
  {\bibfnamefont {K}~\bibnamefont {Eng}}, \bibinfo {author} {\bibfnamefont
  {MP}~\bibnamefont {Jura}}, \bibinfo {author} {\bibfnamefont {AA}~\bibnamefont
  {Kiselev}}, \bibinfo {author} {\bibfnamefont {TD}~\bibnamefont {Ladd}},
  \bibinfo {author} {\bibfnamefont {ST}~\bibnamefont {Merkel}}, \bibinfo
  {author} {\bibfnamefont {I}~\bibnamefont {Milosavljevic}}, \bibinfo {author}
  {\bibfnamefont {EJ}~\bibnamefont {Pritchett}}, \bibinfo {author}
  {\bibfnamefont {MT}~\bibnamefont {Rakher}}, \bibinfo {author} {\bibfnamefont
  {RS}~\bibnamefont {Ross}}, \bibinfo {author} {\bibfnamefont {AE}~\bibnamefont
  {Schmitz}}, \bibinfo {author} {\bibfnamefont {A}~\bibnamefont {Smith}},
  \bibinfo {author} {\bibfnamefont {JA}~\bibnamefont {Wright}}, \bibinfo
  {author} {\bibfnamefont {MF}~\bibnamefont {Gyure}}, \ and\ \bibinfo {author}
  {\bibfnamefont {AT}~\bibnamefont {Hunter}},\ }\bibfield  {title} {\enquote
  {\bibinfo {title} {Reduced sensitivity to charge noise in semiconductor spin
  qubits via symmetric operation},}\ }\href@noop {} {\bibfield  {journal}
  {\bibinfo  {journal} {Physical review letters}\ }\textbf {\bibinfo {volume}
  {116}},\ \bibinfo {pages} {110402} (\bibinfo {year} {2016})}\BibitemShut
  {NoStop}%
\bibitem [{\citenamefont {Cross}\ \emph {et~al.}(2016)\citenamefont {Cross},
  \citenamefont {Magesan}, \citenamefont {Bishop}, \citenamefont {Smolin},\
  and\ \citenamefont {Gambetta}}]{cross2016scalable}%
  \BibitemOpen
  \bibfield  {author} {\bibinfo {author} {\bibfnamefont {Andrew~W}\
  \bibnamefont {Cross}}, \bibinfo {author} {\bibfnamefont {Easwar}\
  \bibnamefont {Magesan}}, \bibinfo {author} {\bibfnamefont {Lev~S}\
  \bibnamefont {Bishop}}, \bibinfo {author} {\bibfnamefont {John~A}\
  \bibnamefont {Smolin}}, \ and\ \bibinfo {author} {\bibfnamefont {Jay~M}\
  \bibnamefont {Gambetta}},\ }\bibfield  {title} {\enquote {\bibinfo {title}
  {Scalable randomised benchmarking of non-clifford gates},}\ }\href@noop {}
  {\bibfield  {journal} {\bibinfo  {journal} {npj Quantum Information}\
  }\textbf {\bibinfo {volume} {2}},\ \bibinfo {pages} {16012} (\bibinfo {year}
  {2016})}\BibitemShut {NoStop}%
\bibitem [{\citenamefont {Hashagen}\ \emph {et~al.}(2018)\citenamefont
  {Hashagen}, \citenamefont {Flammia}, \citenamefont {Gross},\ and\
  \citenamefont {Wallman}}]{Hashagen2018realrandomized}%
  \BibitemOpen
  \bibfield  {author} {\bibinfo {author} {\bibfnamefont {A.~K.}\ \bibnamefont
  {Hashagen}}, \bibinfo {author} {\bibfnamefont {S.~T.}\ \bibnamefont
  {Flammia}}, \bibinfo {author} {\bibfnamefont {D.}~\bibnamefont {Gross}}, \
  and\ \bibinfo {author} {\bibfnamefont {J.~J.}\ \bibnamefont {Wallman}},\
  }\bibfield  {title} {\enquote {\bibinfo {title} {Real {R}andomized
  {B}enchmarking},}\ }\href {\doibase 10.22331/q-2018-08-22-85} {\bibfield
  {journal} {\bibinfo  {journal} {{Quantum}}\ }\textbf {\bibinfo {volume}
  {2}},\ \bibinfo {pages} {85} (\bibinfo {year} {2018})}\BibitemShut {NoStop}%
\end{thebibliography}%

\end{document}